\documentclass[12pt,doublespacing]{article}

\usepackage{fullpage}
\usepackage{setspace}
\usepackage{authblk}

\usepackage{cite}
\usepackage{hyperref} 

\usepackage{graphicx}

\usepackage{amsmath}
\usepackage{amssymb}

\usepackage{subfig}

\begin{document}
\title{Calculation of Photocarrier Generation from Optical Absorption for Time-domain Simulation of Optoelectronic Devices}

\author[1]{Liang Chen}
\author[1]{Ming Dong}
\author[2]{Ran Zhao}
\author[1]{Hakan Bagci\vspace{0.5cm}}

\affil[1]{Liang Chen, Ming Dong, and Hakan Bagci are with the Computer, Electrical, and Mathematical Science and Engineering (CEMSE) Division, King Abdullah University of Science and Technology (KAUST), Thuwal, 23955-6900, Saudi Arabia.
\authorcr e-mails: \{liang.chen, ming.dong, hakan.bagci\}@kaust.edu.sa\vspace{1cm}}

\affil[2]{Ran Zhao was with the Computer, Electrical, and Mathematical Science and Engineering (CEMSE) Division, King Abdullah University of Science and Technology (KAUST), Thuwal 23955-6900, Saudi Arabia.
\authorcr He is now with the School of Electronic Science and Engineering, University of Electronic Science and Technology of China (UESTC), Chengdu 61173, China. 
\authorcr e-mail: ran.zhao@uestc.edu.cn.}

\date{}
\maketitle
\newpage

\begin{abstract}
Photocarrier generation rate in optoelectronic materials is often calculated using the Poynting vector in the frequency domain. However, this approach is not accurate in time-domain simulations of photoconductive devices because the instantaneous Poynting vector does not distinguish between power flux densities of optical and low-frequency electromagnetic fields. The latter is generated by photocurrents and is not supposed to contribute to the photocarrier generation since the corresponding photon energy is smaller than the bandgap energy of the optoelectronic material. This work proposes an optical absorption-based model to accurately calculate the generation rate in time-domain simulations. The proposed approach considers the material dispersion near the optical frequency corresponding to the bandgap energy of the optoelectronic material and calculates the instantaneous optical absorption from the polarization current density associated with this dispersion model. Numerical examples show that the proposed method  is more accurate than the Poynting vector-based approach in calculating the instantaneous optical absorption. The method is further validated against experimental results via simulations of a photoconductive device, where the Poynting vector-based approach results in divergent carrier densities when the low-frequency fields are strong.


\end{abstract}

\section{Introduction}
\label{intro}
Optoelectronic devices are commonly used in electronics industry for various applications~\cite{photonics2019, Chuang2012, Piprek2018, okyay2007, jit2001, zhao2019}. For example, photoconductive devices are often utilized in terahertz (THz) radiation and detection systems, and photovoltaic devices are building blocks of solar cells and photosensors~\cite{Piprek2018, hou2013}. Recent advances in design and fabrication techniques, such as plasmonics-based enhancements~\cite{Lepeshov2017review, Kang2018review, Yu2019plasmon, Yachmenev2019review}, metasurface integration~\cite{Yachmenev2019review, Siday2019}, and nanostructured surface inclusions~\cite{Zhou2018, He2020}, have significantly increased the level of carrier generation that an optoelectronic device can support. This means that numerical schemes, which are indispensable in the design process, must accurately account for the tightly-coupled nonlinear interactions between electromagnetic fields and charge carriers~\cite{Chen2019multiphysics, Chen2020efficient, Chen2021screen}. These numerical schemes solve a coupled system of the Maxwell equations and a carrier transport equation (often the drift-diffusion equation)~\cite{Chuang2012, Piprek2018}. The solution of this system has to be carried out in the time domain due to the presence of strong nonlinearities~\cite{Chen2019multiphysics, Chen2020efficient, Chen2021screen}.

The operation of photoconductive and photovoltaic devices relies on generation of photocarriers upon the absorption of optical fields, which occurs when the photon energy of the field is high enough to excite electrons (typically larger than the bandgap energy of direct-bandgap semiconductors)~\cite{photonics2019, Chuang2012, Piprek2018}. In device simulations, this mechanism is described by a generation rate model that depends on the power flux density of the optical fields~\cite{photonics2019, Chuang2012, Piprek2018}. For simple devices, the generation rate can be estimated using the intensity of the optical incident field, the transmission and the absorption coefficient of the semiconductor~\cite{photonics2019, Chuang2012, Piprek2018, Sano1990, sano1990b, Saeedkia2005, Kirawanich2009, Khiabani2013, Moreno2014}. In complex devices, the optical field distribution is not a simple function of the incident field, and it has to be computed using a full-wave electromagnetic simulator. In this case, the generation rate is obtained using the magnitude of the time-averaged Poynting vector associated with the field distribution. This process is generally executed in the frequency domain~\cite{Neshat2010, Burford2016, Bashirpour2017, Garufo2018, Deinega2012, Sha2012, Li2013solar, Deceglie2012, Fallahpour2015, In2015, Ahn2016, Zhou2018, Khabiri2012} and ignores the coupling from the drift-diffusion equation to the Maxwell equations [interaction of the optical fields with photocarriers and the low-frequency fields generated by the photocurrents (freely moving photocarriers) are ignored].

One cannot directly adopt this Poynting vector-based approach to time-domain simulations where the nonlinear (two-way) coupling between the Maxwell and the drift-diffusion equations is fully accounted for. In this type of simulations, the field distribution includes the optical fields as well as the low-frequency fields that are generated by the photocurrents. These low-frequency fields are strong in the photoconductive devices designed to generate THz radiation from optical fields~\cite{Darrow1992, Benicewicz1994, Kim2006, Loata2007, Chou2013}, but their photon energy, which is proportional to frequency,, is not high enough to excite photocarriers. The corresponding absorptance of an optoelectronic material is high at optical frequencies but negligible at low frequencies~\cite{Blakemore1982, Levinshtein1996, Sehmi2017}. Therefore, the generation rate calculated from the time-dependent Poynting vector, which takes into account the optical as well as the low-frequency fields, is overestimated. In return, this overestimated generation rate yields stronger photocurrents and low-frequency fields. This feedback loop eventually leads to an inaccurate and even divergent/unstable solution.

This work proposes a new approach to calculating the space-time-dependent photocarrier generation rate in optoelectronic materials. This approach considers the material dispersion near the optical frequency corresponding to the bandgap energy of the optoelectronic material. During the time integration, the generation rate is calculated using the instantaneous power dissipation (which is equal to the optical absorption) expressed in terms of this polarization current density associated with the dispersion model~\cite{Cui2004}. The proposed generation rate model is applied to a multiphysics model for optoelectronic devices that is formed by a coupled system of the Maxwell and the drift-diffusion equations.

Numerical examples show that the proposed approach is preferable to the Poynting vector-based method in time-domain simulations. For the purpose of calculating the instantaneous optical absorption, the time-dependent Poynting vector-based method is accurate only when the source is monochromatic and the wave is propagating in a single direction (without scattering), while the proposed method is accurate for general cases. More importantly, in multiphysics simulations of optoelectronic devices, when the two-way couplings between carriers and electromagnetic fields are accounted for, the Poynting vector cannot distinguish the power of optical frequency fields from that of low-frequency radiations, leading to an overestimated carrier generation rate, which results in inaccurate and even divergent carrier densities when the low-frequency radiation is strong. The proposed method is always stable and the simulated device behavior agrees with experimental observations.

The rest of this paper is organized as follows. Section~\ref{formu} describes and formulates the generation rate model, the auxiliary differential equation for the Lorentz dispersion model, and the corresponding time integration scheme. Section~\ref{result} presents numerical examples that validate the accuracy of the proposed method and demonstrate its applicability to photoconductive devices. Furthermore, the reason for the failure of the Poynting vector-based model is analyzed and discussed in this section. Section~\ref{concl} provides a summary of the paper.

\section{Formulation}
\label{formu}
\subsection{Mathematical Model}
Electromagnetic field interactions and charge carrier dynamics on photoconductive and photovoltaic devices are mathematically described by the coupled system of the Maxwell and the drift-diffusion equations as~\cite{Chen2019multiphysics, Chen2020efficient}
\begin{equation}
\label{t_E0}
{\varepsilon_0}{\varepsilon_\infty} \, \partial_t \mathbf{E}(\mathbf{r},t) 
= \nabla \times \mathbf{H}(\mathbf{r},t) - \mathbf{J}_{\mathrm{P}}(\mathbf{r},t) - \mathbf{J}_{\mathrm{D}}(\mathbf{r},t)
\end{equation}
\begin{equation}
\label{t_H0}
{\mu_0}{\mu_r} \, \partial_t \mathbf{H}(\mathbf{r},t) 
= - \nabla \times \mathbf{E}(\mathbf{r},t)
\end{equation}
\begin{equation}
\label{t_N0}
q\partial_t n_c(\mathbf{r},t) 
= \pm \nabla \cdot \mathbf{J}_c(\mathbf{r},t) - q \big[ R(n_e,n_h) - G(\mathbf{E},\mathbf{H}) \big]
\end{equation}
\begin{equation}
\label{t_J0}
\begin{aligned}
{{\bf{J}}_c}({\bf{r}},t) &= q{\mu _c}({\bf{r}})\{ [{\bf{E}}({\bf{r}},t)+ {{\bf{E}}^s}({\bf{r}})]{n_c}({\bf{r}},t) \\ 
	&+ {\bf{E}}({\bf{r}},t)n_c^s({\bf{r}})\} \pm q{d_c}({\bf{r}})\nabla {n_c}({\bf{r}},t).
\end{aligned}
\end{equation}
Here, ${\varepsilon _0}$ and ${\mu _0}$ are the permittivity and the permeability in vacuum, ${\varepsilon _\infty }$ is the permittivity at the infinity frequency, ${\mu _r}$ is the relative permeability, ${\bf{E}}({\bf{r}},t)$ and ${\bf{H}}({\bf{r}},t)$ are the electric and magnetic fields, ${{\bf{J}}_{\mathrm{P}}}({\bf{r}},t)$ is the polarization current density, ${{\bf{J}}_{\mathrm{D}}}({\bf{r}},t)$ is the photocurrent density, ${n_c}({\bf{r}},t)$ is the carrier density with the subscript $c \in \{ e,h\} $ representing the carrier type as $c=e$ for electron and $c=h$ for hole, $R({n_e},{n_h})$ is the recombination rate, $G({\bf{E}},{\bf{H}})$ is the generation rate, ${\mu _c}({\bf{r}})$ and ${d_c}({\bf{r}})$ are the field-dependent mobility and diffusion coefficients~\cite{Chen2020steadystate}, and ${{\bf{E}}^s}({\bf{r}})$ and $n_c^s({\bf{r}})$ are the steady-state electric field and carrier density resulting from the bias voltage and the doping profile~\cite{Chen2020steadystate, Chen2019multiphysics}. In~\eqref{t_E0}, ${{\bf{J}}_{\mathrm{D}}}({\bf{r}},t) = \sum\nolimits_c {{{\bf{J}}_c}({\bf{r}},t)}$, where ${{\bf{J}}_c}({\bf{r}},t)$ is the current density due to the carrier movements and ${{\bf{J}}_{\mathrm{P}}}({\bf{r}},t) = {\partial _t}{{\bf{P}}_{\mathrm{P}}}({\bf{r}},t)$, where ${{\bf{P}}_{\mathrm{P}}}({\bf{r}},t)$ is the polarization density. In~\eqref{t_N0} and~\eqref{t_J0}, the upper and lower signs should be selected for electron ($c = e$) and hole ($c = h$), respectively. In~\eqref{t_J0}, it is assumed that ${{\bf{E}}^s}({\bf{r}})$ and $n_c^s({\bf{r}})$ do not depend on time because the boundary conditions for the Poisson and the stationary drift-diffusion equations (e.g., the Dirichlet boundary conditions on the electrodes) do not change during the transient stage~\cite{Vasileska2010, Chen2020steadystate}. The variation of the electromagnetic fields in time due to the time-dependent ${{\bf{J}}_{\mathrm{D}}}({\bf{r}},t)$ and ${{\bf{J}}_{\mathrm{P}}}({\bf{r}},t)$ is fully captured in ${\bf{E}}({\bf{r}},t)$ and ${\bf{H}}({\bf{r}},t)$ by solving the Maxwell equations in~\eqref{t_E0} and~\eqref{t_H0}~\cite{Chen2020efficient}.

In~\eqref{t_N0}, $G({\bf{E}},{\bf{H}})$ is the rate of photocarrier generation upon the absorption of the optical field energy and is expressed as~\cite{photonics2019, Chuang2012, Piprek2018}
\begin{equation}
\label{generation}
G({\bf{E}},{\bf{H}}) = \eta \frac{{{P^{{\mathrm{abs}}}}({\bf{r}},t)}}{{{E^{{\mathrm{ph}}}}}}.
\end{equation}
Here, $\eta $ is the intrinsic quantum efficiency, i.e., the number of electron-hole pairs generated for every absorbed photon, ${P^{{\mathrm{abs}}}}({\bf{r}},t)$ is the absorbed power density of the optical field, ${E^{{\mathrm{ph}}}} = h\nu $ is the photon energy, $h$ is the Planck constant, and $\nu $ is the frequency. Note that, $\nu $ must be high enough so that ${E^{{\mathrm{ph}}}}$ can excite electrons (e.g., usually, ${E^{{\mathrm{ph}}}}$ should be larger than the bandgap energy in direct-bandgap semiconductors)~\cite{photonics2019, Chuang2012, Piprek2018}.

In the literature, the coupled system described by~\eqref{t_E0}-\eqref{generation} has often been simplified under certain approximations and assumptions. For very simple devices, where the electromagnetic interactions can be approximated using only the incident field, ${P^{{\mathrm{abs}}}}({\bf{r}},t)$ in~\eqref{generation} is reduced to~\cite{Sano1990, sano1990b, Saeedkia2005, Kirawanich2009, Chuang2012, Khiabani2013, Moreno2014, Piprek2018, photonics2019}
\begin{equation}
\label{PabsA}
{P^{{\mathrm{abs}}}}({\bf{r}},t) = {P_0}T\alpha {e^{ - \alpha d}}f({\bf{r}},t).
\end{equation}
Here, ${P_0}$ is the peak power flux density of the incident optical pulse, $T$ is the transmittance at the air-semiconductor interface, $\alpha$ is the absorption coefficient (the imaginary part of the permittivity is sometimes used instead~\cite{Burford2016, Bashirpour2017}), $d$ is the penetration depth, and $f({\bf{r}},t)$ accounts for the spatial distribution and the temporal delay of the optical pulse. Note that~\eqref{PabsA} is obtained under the assumption that the incident optical pulse enters the semiconductor layer through a simple air-semiconductor interface. It is also assumed that $f({\bf{r}},t)$ is narrowband. This assumption permits a constant value to be used for $\alpha$, which in reality is frequency-dependent~\cite{Blakemore1982, Levinshtein1996, Sehmi2017}.

For more complex devices, such as those that are integrated with nanostructures and metasurfaces~\cite{Lepeshov2017review, Kang2018review, Yu2019plasmon, Yachmenev2019review, Siday2019, Zhou2018, He2020}, the simplified expression in~\eqref{PabsA} cannot be used to estimate $G({\bf{E}},{\bf{H}})$. Some of the methods that have been developed to overcome this bottleneck solve a simplified version of the coupled system in~\eqref{t_E0}-\eqref{t_J0}~\cite{Neshat2010, Burford2016, Bashirpour2017, Garufo2018, Deinega2012, Sha2012, Li2013solar, Deceglie2012, Fallahpour2015, In2015, Ahn2016, Zhou2018, Khabiri2012}. These methods ignore ${{\bf{J}}_{\mathrm{D}}}({\bf{r}},t)$ in~\eqref{t_E0}, i.e., they do not model the coupling from the drift-diffusion equations to the Maxwell equations, and solve the Maxwell equations~\eqref{t_E0}-\eqref{t_H0} in the frequency domain for ${\bf{\bar E}}({\bf{r}},{\nu _0})$ and ${\bf{\bar H}}({\bf{r}},{\nu _0})$ [Fourier transforms of ${\bf{E}}({\bf{r}},t)$ and ${\bf{H}}({\bf{r}},t)$ at frequency ${\nu _0}$]. Then, the time-averaged Poynting vector ${\bf{\bar S}}({\bf{r}},{\nu _0})$ is used to calculate ${P^{{\mathrm{abs}}}}({\bf{r}},t)$~\cite{Neshat2010, Burford2016, Bashirpour2017, Garufo2018, Deinega2012, Sha2012, Li2013solar, Deceglie2012, Fallahpour2015, In2015, Ahn2016, Zhou2018, Khabiri2012}:
\begin{equation}
\label{PabsS}
{P^{{\mathrm{abs}}}}({\bf{r}},t) = \alpha |{\bf{\bar S}}({\bf{r}},{\nu _0})|f(t).
\end{equation}
Here, ${\bf{\bar S}}({\bf{r}},{\nu _0}) = \operatorname{Re} \{ {\bf{\bar E}}({\bf{r}},{\nu _0}) \times {{\bf{\bar H}}^*}({\bf{r}},{\nu _0})\} /2$, operator $\operatorname{Re} \{ .\} $ and superscript ``$*$'' denote the real part and the complex conjugate, respectively, and $f(t)$ accounts for the envelope of the incident field in time~\cite{Chuang2012, Khabiri2012, Burford2016, Bashirpour2017}. In~\cite{Khabiri2012}, $ - \nabla \cdot {\bf{\bar S}}({\bf{r}},{\nu _0})$ is used instead of $\alpha |{\bf{\bar S}}({\bf{r}},{\nu _0})|$. Because ${\bf{\bar S}}({\bf{r}},{\nu _0})$ is defined in the frequency domain, $f(t)$ has to be a slowly varying function (compared to ${\nu _0}$). And this means that ${\bf{\bar S}}({\bf{r}},\nu )$ at all frequencies $\nu$ in the band of $f(t)$ are approximated by ${\bf{\bar S}}({\bf{r}},{\nu _0})$. Note that ${\nu _0}$ is chosen as the center frequency of $f(t)$'s band. For photovoltaic devices, usually a wide frequency band is considered, and ${P^{{\mathrm{abs}}}}({\bf{r}},t)$ is calculated at each sampling frequency, with $f(t) = 1$ and weighted by the solar radiation spectrum~\cite{Deinega2012, Sha2012, Li2013solar, Deceglie2012, Fallahpour2015, In2015, Ahn2016, Zhou2018}. After ${P^{{\mathrm{abs}}}}({\bf{r}},t)$ is calculated using~\eqref{PabsS}, $G({\bf{E}},{\bf{H}})$ is calculated using~\eqref{generation}. Then, a modified version of~\eqref{t_N0}-\eqref{t_J0}, where $[{\bf{E}}({\bf{r}},t) + {{\bf{E}}^s}({\bf{r}})]$ is replaced by a single electric field variable, is solved together with the Poisson equation~\cite{Khabiri2012, Burford2016, Bashirpour2017}. Since this two-step approach to solving the simplified version of~\eqref{t_E0}-\eqref{generation} ignores the influence of the moving carriers on the optical fields [i.e., ignores ${{\bf{J}}_{\mathrm{D}}}({\bf{r}},t)$ in~\eqref{t_E0}], it fails to capture several saturation effects that are observed in experiments especially when the carrier density is high~\cite{Darrow1992, Chen2021screen}.

These saturation effects result from the coupling between the photocarriers and the electromagnetic fields. High carrier density levels result in a large effective photoconductivity, which in return blocks the optical fields from entering/penetrating the active region of the device~\cite{Darrow1992, Chen2020efficient, Chen2021screen}. If ${{\bf{J}}_{\mathrm{D}}}({\bf{r}},t)$ is ignored in~\eqref{t_E0}, the effect of the photoconductivity on ${\bf{E}}({\bf{r}},t)$ and ${\bf{H}}({\bf{r}},t)$ cannot be accounted for, which consequently means that saturation effects are not observed in the simulation results. A more accurate approach, which can account for the saturation effects, is to directly solve~\eqref{t_E0}-\eqref{generation} in the time domain without ignoring any coupling terms~\cite{Chen2019multiphysics, Chen2020efficient, Chen2021screen}. For this approach, one could choose to use the time-dependent Poynting vector ${\bf{S}}({\bf{r}},t)$ to calculate ${P^{{\mathrm{abs}}}}({\bf{r}},t)$ directly in the time domain using
\begin{equation}
\label{PabsS1}
{P^{{\mathrm{abs}}}}({\bf{r}},t) = \alpha |{\bf{S}}({\bf{r}},t)|
\end{equation}
where ${\bf{S}}({\bf{r}},t) = {\bf{E}}({\bf{r}},t) \times {\bf{H}}({\bf{r}},t)$. However, the main problem with using~\eqref{PabsS1} to calculate $G({\bf{E}},{\bf{H}})$ is that ${\bf{S}}({\bf{r}},t)$ represents the power flux density of the electromagnetic fields at all frequencies, including those at the low frequencies (e.g., THz frequencies in photoconductive devices), which are generated by the photocurrents. However, at the low frequencies, ${E^{{\mathrm{ph}}}}$ is smaller than the bandgap energy and the power flux density of the low-frequency fields do not contribute to the generation of the photocarriers and should not be included in the calculation of $G({\bf{E}},{\bf{H}})$.

To overcome this problem in the time-domain solution of the fully coupled system in~\eqref{t_E0}-\eqref{t_J0}, a new model to calculate ${P^{{\mathrm{abs}}}}({\bf{r}},t)$ is developed as described next. Consider the Poynting theorem for the electromagnetic system represented by~\eqref{t_E0}-\eqref{t_J0}~\cite{Haus1989book}
\begin{equation}
\label{PEJ}
\nabla \cdot {\bf{S}}({\bf{r}},t) + {\partial _t}W({\bf{r}},t) + {P_{\mathrm{D}}}({\bf{r}},t) + {P_{\mathrm{P}}}({\bf{r}},t) = 0
\end{equation}
where $W({\bf{r}},t) = ({\varepsilon _0}{\varepsilon _\infty }|{\bf{E}}({\bf{r}},t){|^2} + {\mu _0}{\mu _r}|{\bf{H}}({\bf{r}},t){|^2})/2$ is the sum of the stored electric and magnetic energy densities, which are due to the non-dispersive polarization and the magnetization processes, and ${P_{\mathrm{D}}}({\bf{r}},t) = {\bf{E}}({\bf{r}},t) \cdot {{\bf{J}}_{\mathrm{D}}}({\bf{r}},t)$ and ${P_{\mathrm{P}}}({\bf{r}},t) = {\bf{E}}({\bf{r}},t) \cdot {{\bf{J}}_{\mathrm{P}}}({\bf{r}},t)$ are the power densities associated with the conduction and the dispersive polarization current densities, respectively. ${P_{\mathrm{D}}}({\bf{r}},t)$ represents the conduction power loss~\cite{Haus1989book}, where ${{\bf{J}}_{\mathrm{D}}}({\bf{r}},t)$ is calculated using the drift-diffusion equation (instead of the Ohm law in a conductive medium). ${P_{\mathrm{P}}}({\bf{r}},t)$ represents the energy storage and the dissipation in the polarization process. The imaginary part of the permittivity corresponds to power dissipation, which is referred as the optical absorption (when the imaginary part of the permittivity has a positive value).

To calculate $G({\bf{E}},{\bf{H}})$ from the optical absorption represented by ${P^{{\mathrm{abs}}}}({\bf{r}},t)$, it is essential to separate the power dissipation from the energy storage in ${P_{\mathrm{P}}}({\bf{r}},t)$. To this end, it is assumed that the permittivity $\varepsilon (\omega )$ is represented using a multipole Lorentz model with poles residing in the frequency range of interest:
\begin{equation}
\label{Lorentz} 
\varepsilon (\omega ) =  {\varepsilon _0}{\varepsilon _\infty } + {\varepsilon _0}\sum\limits_{n = 1}^N {\frac{{\omega _{{\mathrm{p}},n}^2}}{{\omega _{{\mathrm{o}},n}^2 - {\omega ^2} - i{\gamma _n}\omega }}}.
\end{equation}
Here, $\omega=2\pi \nu$ is angular frequency, ${\omega _{{\mathrm{o}},n}}$ is the angular resonance frequency, ${\omega _{{\mathrm{p}},n}}$ is the angular plasma frequency, ${\gamma _n}$ is the damping constant, and $N$ is the number of poles. The electric flux density corresponding to ${{\bf{P}}_{\mathrm{P}}}({\bf{r}},t)$ is expressed as ${\bf{D}}({\bf{r}},t)\! =\! {\varepsilon _0}{\varepsilon _\infty }{\bf{E}}({\bf{r}},t) + {{\bf{P}}_{\mathrm{P}}}({\bf{r}},t)$, where ${{\bf{P}}_{\mathrm{P}}}({\bf{r}},t)\! =\! \sum\nolimits_{n = 1}^N {{{\bf{P}}_n}({\bf{r}},t)} $. Inserting~\eqref{Lorentz} into ${\bf{\bar D}}({\bf{r}},\omega )\! =\! \varepsilon (\omega ){\bf{\bar E}}({\bf{r}},\omega )$, where ${\bf{\bar D}}({\bf{r}},\omega )$ and ${\bf{\bar E}}({\bf{r}},\omega )$ are Fourier transforms of ${\bf{D}}({\bf{r}},t)$ and ${\bf{E}}({\bf{r}},t)$, and converting the resulting equation into the time domain, one can see that ${{\bf{P}}_n}({\bf{r}},t)$ satisfies
\begin{equation}
\label{Lorentz1}
\partial _t^2{{\mathbf{P}}_n}({\mathbf{r}},t) + {\gamma _n}{\partial _t}{{\mathbf{P}}_n}({\mathbf{r}},t) + \omega _{{\mathrm{o}},n}^2{{\mathbf{P}}_n}({\mathbf{r}},t)= {\varepsilon _0}\omega _{{\mathrm{p}},n}^2{\mathbf{E}}({\mathbf{r}},t).
\end{equation}
Let ${{\bf{J}}_{\mathrm{P}}}({\bf{r}},t)\! =\! \sum\nolimits_{n = 1}^N {{{\bf{J}}_n}({\bf{r}},t)}$, ${{\bf{J}}_n}({\bf{r}},t)\! =\! {\partial _t}{{\bf{P}}_n}({\bf{r}},t)$. Inserting this summation and the expression for ${\bf{E}}({\bf{r}},t)$ from~\eqref{Lorentz1} into ${P_{\mathrm{P}}}({\bf{r}},t)\! =\! {\bf{E}}({\bf{r}},t) \cdot {{\bf{J}}_{\mathrm{P}}}({\bf{r}},t)$ yields
\begin{equation}
\label{Summation}
    P_{\mathrm{P}}(\mathbf{r}, t) = \sum_{n=1}^N P_{{\mathrm{P}},n}(\mathbf{r}, t)
= \sum_{n=1}^N \left\{
  \frac{1}{2 \varepsilon_0 \omega_{{\mathrm{p}},n}^2} 
  \frac{\partial}{\partial t}
  \left(\left| \mathbf{J}_n \right|^2 +
    \omega_{{\mathrm{o}},n}^2 \left| \mathbf{P}_n \right|^2
  \right)+\frac{\gamma_n}{\varepsilon_0 \omega_{{\mathrm{p}},n}^2}
  \left| \mathbf{J}_n \right|^2\right\}.
\end{equation}
Here, the first-order time-derivative term is the rate of change of the energy storage that can be combined into ${\partial _t}W({\bf{r}},t)$ in~\eqref{PEJ}, and the second term, being positive and proportional to ${\gamma _n}$, is the power dissipation~\cite{Cui2004, Haus1989book, Smith1998, Huang2015}. Consequently, the optical absorption ${P^{{\mathrm{abs}}}}({\bf{r}},t)$ is given by the summation of the second term over all poles:
\begin{equation}
\label{Pabs1} 
{P^{{\mathrm{abs}}}}({\bf{r}},t) = \sum\limits_{n = 1}^N {P_n^{{\mathrm{abs}}}({\bf{r}},t)} = \sum\limits_{n = 1}^N {\frac{{{\gamma _n}}}{{{\varepsilon _0}\omega _{{\mathrm{p}},n}^2}}|{{\bf{J}}_n}({\bf{r}},t){|^2}}.
\end{equation}
Similarly, $G({\bf{E}},{\bf{H}})$ can be written as a summation and calculated as such:
\begin{equation}
\label{generation1}
G({\bf{E}},{\bf{H}}) = \sum\limits_{n = 1}^N {{G_n}({\bf{E}},{\bf{H}})} = \sum\limits_{n = 1}^N {\eta \frac{{P_n^{{\mathrm{abs}}}({\bf{r}},t)}}{{{E^{{\mathrm{ph}}}}}}}.
\end{equation}
\subsection{Time Integration}
The auxiliary differential equation methods developed to incorporate the Lorentz model into the time-domain numerical schemes has been studied well in the literature (\cite{Taflove2005, Gedney2012} and references therein). In what follows, a modified auxiliary equation method, which uses ${{\bf{J}}_n}({\bf{r}},t)$ as the auxiliary variable to directly calculate the optical absorption, is formulated. First,~\eqref{Lorentz1} is rewritten as
\begin{equation}
\label{Lorentz2}
\partial_t \mathbf{P}_n(\mathbf{r}, t) = \mathbf{J}_n(\mathbf{r}, t)
\end{equation}
\begin{equation}
\label{Lorentz3}
\partial_t \mathbf{J}_n(\mathbf{r}, t) 
+ \gamma_n \mathbf{J}_n(\mathbf{r}, t) + \omega_{\mathrm{o},n}^2 \mathbf{P}_n(\mathbf{r}, t)= \varepsilon_0 \omega_{\mathrm{p},n}^2 \mathbf{E}(\mathbf{r}, t)
\end{equation}
Equations~\eqref{t_E0}-\eqref{t_J0} and~\eqref{generation1}-\eqref{Lorentz3} form the final system to be integrated in time. Due to the time-scale difference, the Maxwell equations~\eqref{t_E0}-\eqref{t_H0} and~\eqref{Lorentz2}-\eqref{Lorentz3} and the drift-diffusion equation~\eqref{t_N0}-\eqref{t_J0} are updated separately with independent but coupled schemes~\cite{Chen2019multiphysics}. The low-storage five-stage fourth-order Runge-Kutta time integration scheme~\cite{Hesthaven2008} is used for~\eqref{t_E0}-\eqref{t_H0} and~\eqref{Lorentz2}-\eqref{Lorentz3}:
\begin{flalign*}
	&{{\bf{H}}^{(0)}} = {\bf{H}}({\bf{r}},k\Delta t)&\\
    &{{\bf{E}}^{(0)}} = {\bf{E}}({\bf{r}},k\Delta t)& \\
	& {\bf{J}}_n^{(0)} = {{\bf{J}}_n}({\bf{r}},k\Delta t), \,n = 1,\ldots,N & \\
    &{\bf{P}}_n^{(0)} = {{\bf{P}}_n}({\bf{r}},k\Delta t),\,n = 1,\ldots,N &\\
	& {\bf{E}}_{{\mathrm{res}}}^{(0)} = 0&\\
    &{\bf{H}}_{{\mathrm{res}}}^{(0)} = 0 &\\
    &{\bf{P}}_{n,{\mathrm{res}}}^{(0)} = 0&\\
    &{\bf{J}}_{n,{\mathrm{res}}}^{(0)} = 0&\\
    & {\mathrm{for}}\,i = 1,\ldots,5 &\\
	& \quad {\bf{E}}_{{\mathrm{rhs}}}^{(i)} = [\nabla \times {{\bf{H}}^{(i - 1)}} - {{\bf{J}}_{\mathrm{D}}} - \sum\nolimits_{n = 1}^N {{\bf{J}}_n^{(i - 1)}} ]/({\varepsilon _0}{\varepsilon _\infty }) &\\ 
	& \quad {\bf{E}}_{{\mathrm{res}}}^{(i)} = {A^{(i)}}{\bf{E}}_{{\mathrm{res}}}^{(i - 1)} + \Delta t{\bf{E}}_{{\mathrm{rhs}}}^{(i)} & \\
	& \quad {\bf{H}}_{{\mathrm{res}}}^{(i)} = {A^{(i)}}{\bf{H}}_{{\mathrm{res}}}^{(i - 1)} - \Delta t\nabla \times {{\bf{E}}^{(i - 1)}}/{\mu _0} &
    \end{flalign*}
    \begin{flalign*}
	& \quad {\mathrm{for}}\,n = 1,\dots,N & \\ 
	& \quad \quad {\bf{P}}_{n,{\mathrm{res}}}^{(i)} = {A^{(i)}}{\bf{P}}_{n,{\mathrm{res}}}^{(i-1)} + \Delta t{\bf{J}}_n^{(i-1)} &\\ 
	& \quad \quad {\bf{J}}_{n,{\mathrm{res}}}^{(i)} = {A^{(i)}}{\bf{J}}_{n,{\mathrm{res}}}^{(i-1)} +\Delta t[{\varepsilon _0}\omega _{{\mathrm{p}},n}^2\!{{\bf{E}}^{(i-1)}}  - \omega _{{\mathrm{o}},n}^2{\bf{P}}_n^{(i-1)}- {\gamma _n}{\bf{J}}_n^{(i-1)}] &\\ 
	& \quad {\mathrm{end\,for}}&\\
    & \quad {{\bf{E}}^{(i)}} = {{\bf{E}}^{(i-1)}} + {B^{(i)}}{\bf{E}}_{{\mathrm{res}}}^{(i)} &\\
	& \quad {{\bf{H}}^{(i)}} = {{\bf{H}}^{(i-1)}} + {B^{(i)}}{\bf{H}}_{{\mathrm{res}}}^{(i)}&\\
	& \quad {\mathrm{for}}\,n = 1,\dots,N &\\ 
	& \quad \quad {\bf{P}}_n^{(i)} = {\bf{P}}_n^{(i-1)} + {B^{(i)}}{\bf{P}}_{n,{\mathrm{res}}}^{(i)} &\\ 
	& \quad \quad {\bf{J}}_n^{(i)} = {\bf{J}}_n^{(i-1)} + {B^{(i)}}{\bf{J}}_{n,{\mathrm{res}}}^{(i)} &\\ 
	& \quad {\mathrm{end\,for}} &\\ 
	& {\mathrm{end\,for}} &\\ 
	& {\bf{E}}({\bf{r}},[k + 1]\Delta t)={{\bf{E}}^{(5)}}&\\ 
    &{\bf{H}}({\bf{r}},[k +1]\Delta t)={{\bf{H}}^{(5)}} &\\ 
	& {{\bf{J}}_n}({\bf{r}},[k+1]\Delta t)={\bf{J}}_n^{(5)},\,n=1,\ldots, N,&\\
    &{{\bf{P}}_n}({\bf{r}},[k+1]\Delta t)={\bf{P}}_n^{(5)},\, n=1,\ldots, N.&
\end{flalign*}
Here, $\Delta t$ is the time-step size, ${A^{(i)}}$ and ${B^{(i)}}$ are Runge-Kutta coefficients, and superscript $(i)$ refers to the variables used/updated in $i{\mathrm{th}}$ stage of the Runge-Kutta scheme. In the above algorithm, the update from time step $k$ to time step $(k\! +\! 1)$ is demonstrated and it is assumed that ${{\bf{J}}_{\mathrm{D}}}$ is computed from the time-integration updates of the drift-diffusion equation. Note that the space-dependence of the intermediary variables is ignored for the sake of simplicity in the presentation and it is assumed the spatial discretization is properly done, for example using a discontinuous Galerkin scheme~\cite{Chen2019multiphysics,Chen2020steadystate,zhu2025} or a finite volume-finite element method~\cite{wang2021}.

At the end of Runge-Kutta updates, ${{\bf{J}}_n}({\bf{r}},[k\! +\! 1]\Delta t)$ is used in~\eqref{Pabs1} to compute ${P^{{\mathrm{abs}}}}({\bf{r}},[k\! +\! 1]\Delta t)$, and ${P^{{\mathrm{abs}}}}({\bf{r}},[k\! +\! 1]\Delta t)$ is used in~\eqref{generation1} to compute ${\left. {G({\bf{E}},{\bf{H}})} \right|_{t = [k\! +\! 1]\Delta t}}$. Then, ${\left. {G({\bf{E}},{\bf{H}})} \right|_{t = [k\! +\! 1]\Delta t}}$ is used in the time-integration scheme for the drift-diffusion equation~\eqref{t_N0}-\eqref{t_J0}. In this work, this is done using a third-order total-variation-diminishing Runge-Kutta scheme~\cite{Shu1988}. Note that ${{\bf{J}}_{\mathrm{D}}}({\bf{r}},t)$ varies much slower than the optical fields, and therefore, the time-step size for the drift-diffusion equation can be much larger~\cite{Chen2019multiphysics}. In this case, several ${\left. {G({\bf{E}},{\bf{H}})} \right|_{t = k\Delta t}}$ with $(k\Delta t)$s that fall within this time-step are averaged to be used in the time-integration of the drift-diffusion equation. 

\subsection{Comments}
${E^{{\mathrm{ph}}}}$ in~\eqref{generation} and~\eqref{generation1} is frequency-dependent and therefore these two equations cannot be directly used to calculate the photon flux of a wideband optical pulse. This is not a problem for simulations of photoconductive devices because the source is narrowband (less than $1\% $) with center frequency ${\nu _0}$ and $h{\nu _0}$ is large enough to excite electrons. Then~\eqref{generation} can be used to calculate the photon flux.
For simulations of photovoltaic devices, the frequency range of interest usually covers the entire visible spectrum. Like the frequency-domain methods, multiple simulations with different narrowband sources can be executed to cover the entire frequency range. In this case, the proposed method might reduce the number of simulations since it allows for a wideband source and a dispersion model with multiple poles, each of which covers a narrow band. Other poles or dispersion models in other frequency ranges can also be included; however, only those poles contributing to the optical absorption (e.g., those with corresponding photon energies larger than the bandgap energy) should be included in~\eqref{generation1}. The proposed method relies on the separation of the dissipated power density from the reactive power density in~\eqref{Summation}. This can also be done for other dispersion models~\cite{Cui2004, Haus1989book, Smith1998, Huang2015}

It should be noted that, in the above formulation, the Lorentz dispersion model is used to phenomenally describe the interband electron transition upon the absorption of optical waves. The corresponding polarization current density $\mathbf{J}_{\mathrm{P}}(\mathbf{r},t)$ is associated with the electrical displacement of bound charges. On the other hand, the freely-moving photocarriers are modeled with the drift-diffusion model and the corresponding conductivity enters the Maxwell equations through the photocurrent density $\mathbf{J}_{\mathrm{D}}(\mathbf{r},t)$. Hence, the Lorentz model parameters are assumed independent of the density of freely-moving carriers. Also note that, in this work, the Maxwell-drift-diffusion system is used to illustrate the carrier generation rate model. But the generation rate model does not rely on the drift-diffusion model and is applicable to other carrier transport models for semiconductor devices, such as the hydrodynamic model~\cite{Vasileska2010}.

It should be noted that in the Poynting vector-based method, the absorption coefficient in~\eqref{PabsS1} is ambiguous when the source is non-monochromatic since the single-valued ${\bf{S}}({{\bf{r}}},t)$ corresponds to all frequency components. State-of-the-art time-domain solvers (see, e.g., Dassault Syst\`emes CST Studio Suite, Ansys Lumerical FDTD, MEEP\cite{MEEP}, and EMTL\cite{Deinega2012, Bhattacharya2019}, etc.) calculate the Fourier transform of ${\bf{S}}({{\bf{r}}},t)$ after all time iterations are finished and consider different absorption coefficients for different frequency components. This is essentially the same as using a frequency-domain method to calculate the time-averaged Poynting vector at different frequencies (with proper weights corresponding to the source power spectrum)~\cite{Shang2016, Anderson2020}. However, as discussed in Section~\ref{intro}, to account for the nonlinear coupling between the photocarriers and the electromagnetic fields, the instantaneous absorbed power density (and carrier generation rate) has to be calculated during the time integration since the generated carriers will affect the electromagnetic field propagation (and the Poynting vector distribution) in subsequent time iterations.

\section{Numerical Results}
\label{result}

\subsection{Validation}
\label{validation}
\begin{figure}[ht!]
	\centering
    \begin{tabular}[b]{c}
        \includegraphics[width=0.6\columnwidth]{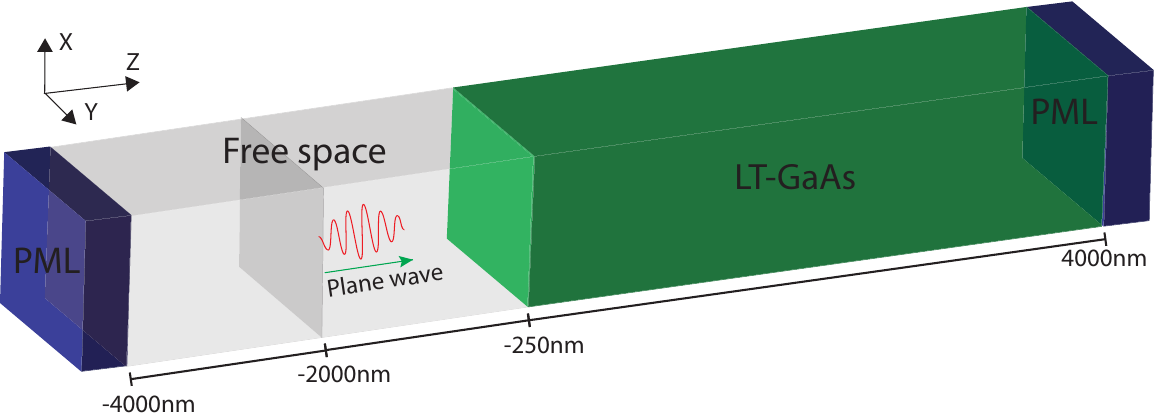}\\
        \footnotesize (a)
    \end{tabular}
    
    \begin{tabular}[b]{c}
         \includegraphics[width=0.45\columnwidth]{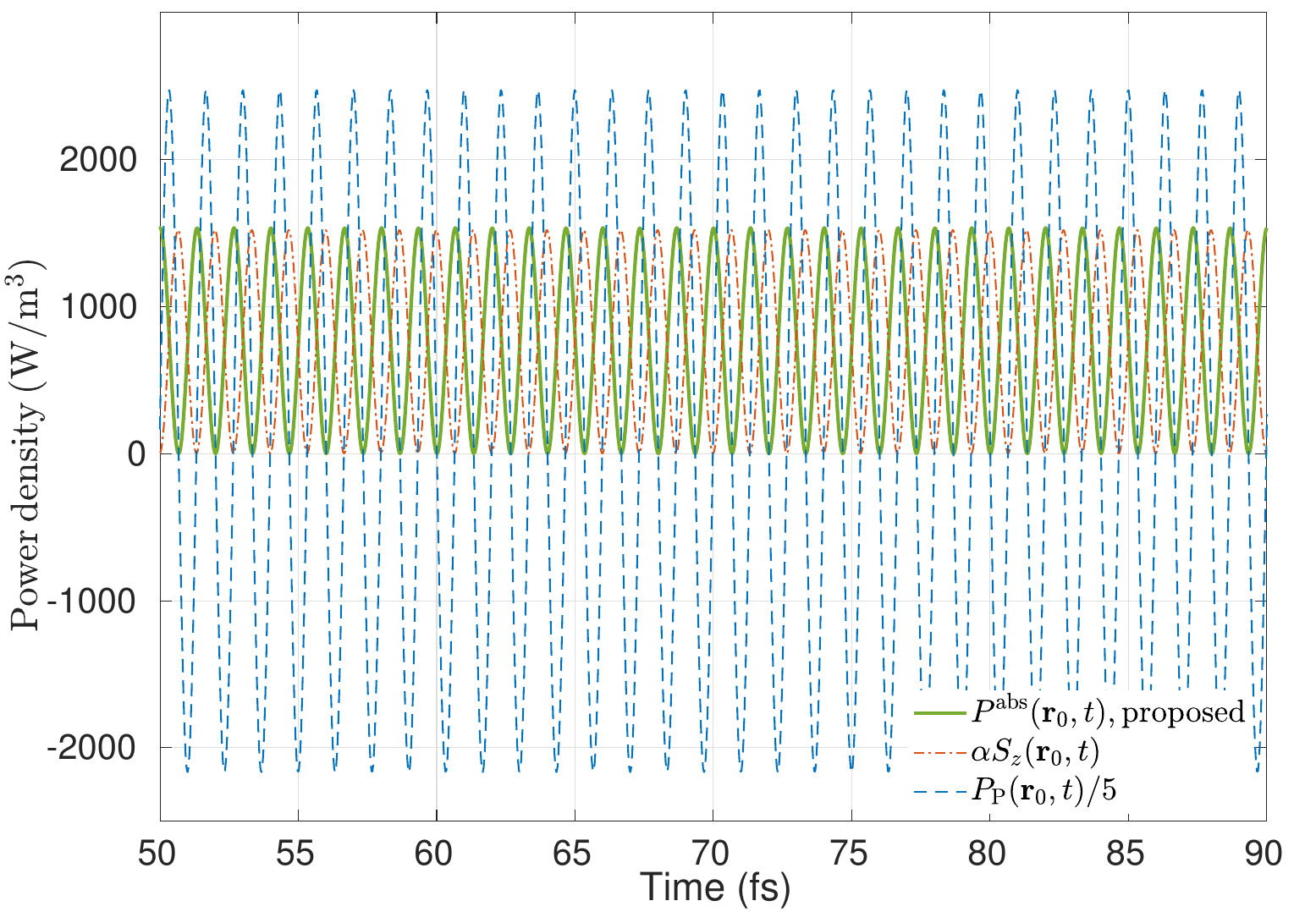}\\
         \footnotesize (b)
    \end{tabular}
    
    \begin{tabular}[b]{c}
         \includegraphics[width=0.45\columnwidth]{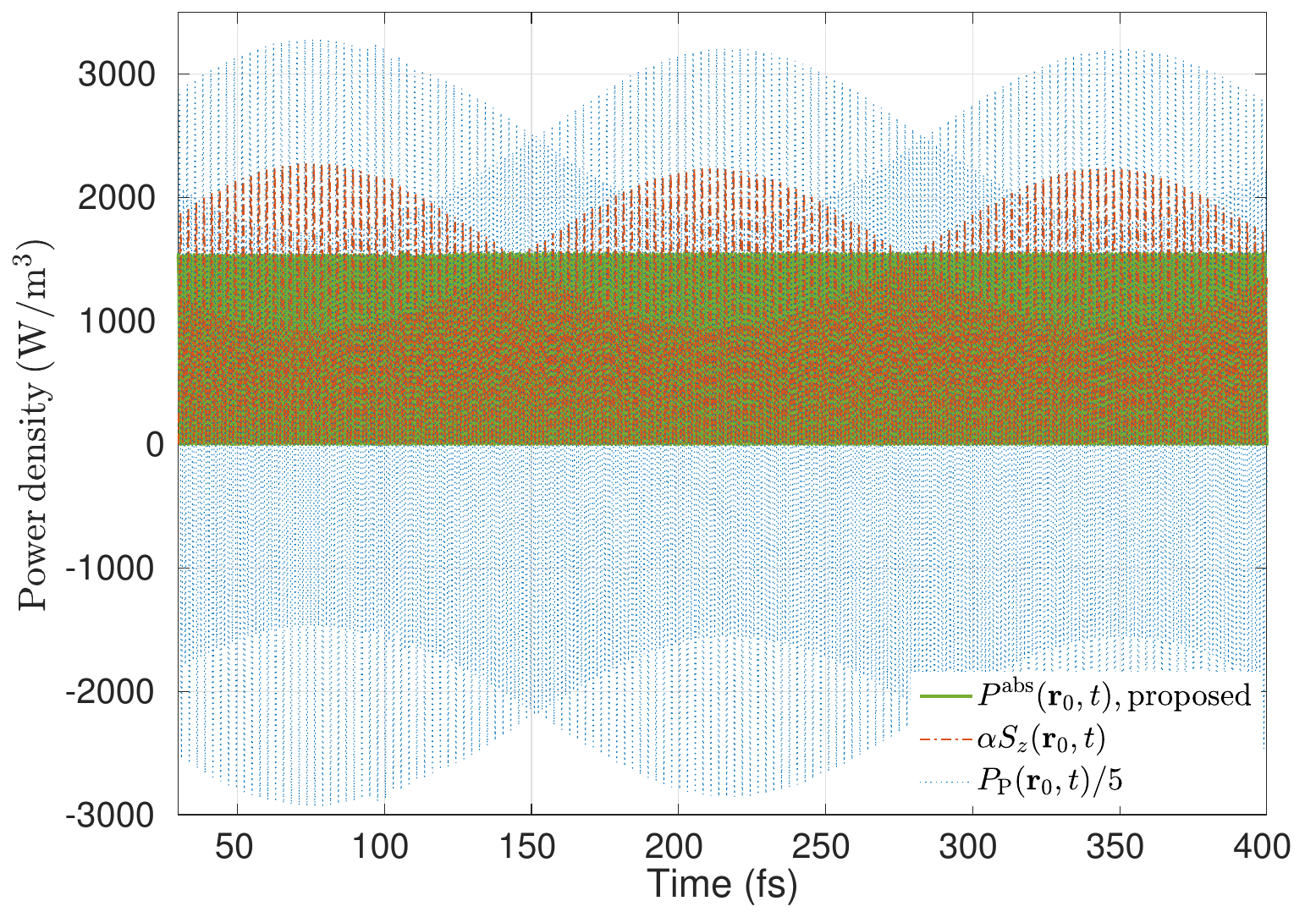}\\
         \footnotesize (c)
    \end{tabular}
	\caption{Simulation with LT-GaAs filling $z\ge -250~\mathrm{nm}$. (a) Geometry configuration. ${P^{{\mathrm{abs}}}}({{\bf{r}}_0},t)$ [calculated using~\eqref{Pabs1}], $\alpha {S_z}({{\bf{r}}_0},t)$ and ${P_{\mathrm{P}}}({{\bf{r}}_0},t)$ at ${{\bf{r}}_0} = (0,0,0)$ (b) under a monochromatic excitation and (c) under the excitation with a modulated signal $sin(2\pi\nu t)+0.2sin(0.02\pi \nu t)$. Note that ${P_{\mathrm{P}}}({{\bf{r}}_0},t)$ is scaled for a better demonstration.}
	\label{abs0}
\end{figure}

In this subsection, the proposed method for calculating the space-time-dependent optical absorption ${P^{{\mathrm{abs}}}}({\bf{r}},t)$ is validated. Consider the space with a single interface. The optoelectronic material low-temperature-grown GaAs (LT-GaAs) fills the half-space $z\geq -250~\mathrm{nm}$ and the remaining half-space is free space. Fig.~\ref{abs0}(a) depicts the simulation setup. Since the calculation of ${P^{{\mathrm{abs}}}}({\bf{r}},t)$ does not require the solution of the drift-diffusion equation, only optical properties of LT-GaAs are considered in this simulation. Periodic boundary conditions are used along the $x$ and $y$ directions. Perfectly matched layers~\cite{Chen2020pml, Gedney2009} are used along the $z$ direction. The Lorentz model is generated via a ``fitting'' scheme applied to the experimentally measured permittivity of LT-GaAs in the frequency range $[0,600\;{\mathrm{THz}}]$~\cite{Blakemore1982}. A single Lorentz pole with parameters ${\varepsilon _\infty} = 5.79$, ${\omega _{\mathrm{o}}} = 4.67 \times {10^{15}}$, ${\omega _{\mathrm{p}}} = 1.06 \times {10^{16}}$, $\gamma = 4.56 \times {10^{14}}$ yields maximum relative errors of $0.49\% $ and $0.66\%$ with respect to the experimental values in the frequency range $[0,600\,{\mathrm{THz}}]$ for the real and the imaginary parts of the permittivity, respectively. All materials are nonmagnetic.

A monochromatic $x$-polarized plane wave with frequency $\nu = 375\,{\mathrm{THz}}$ is normally incident on the LT-GaAs interface from the free-space. The complex relative permittivity at the frequency of excitation is $12.69 + 0.457i$. The corresponding absorption coefficient $\alpha = 1.01 \times {10^6}\,{{\mathrm{m}}^{ - 1}}$. Three methods for calculating the space-time-dependent absorbed power density are compared, i.e., (I) ${P^{{\mathrm{abs}}}}({{\bf{r}}},t)$ calculated using the proposed method [via~\eqref{Pabs1}], (II) $\alpha |\bf{S}({{\bf{r}}},t)|$ [corresponding to~\eqref{PabsS1}], and (III) ${P_{\mathrm{P}}}({{\bf{r}}},t) = {\bf{E}}({{\bf{r}}},t) \cdot  {{\bf{J}}_{\mathrm{P}}}({{\bf{r}}},t)$. 

Note that  ${P_{\mathrm{P}}}({{\bf{r}}},t)$ contains not only the absorbed power density but also the reactive power density. However, since it is directly defined from the Poynting theorem [see~\eqref{PEJ}] and the time average of the reactive power is zero, in the following, the time-averaged ${P_{\mathrm{P}}}({{\bf{r}}},t)$ is used as the reference to test the time-averaged absorbed power density.

Fig.~\ref{abs0} (b) shows ${P^{{\mathrm{abs}}}}({{\bf{r}}_0},t)$ calculated using the proposed method, $\alpha {S_z}({{\bf{r}}_0},t)$, and ${P_{\mathrm{P}}}({{\bf{r}}_0},t)$ at ${{\bf{r}}_0} = (0,0,0)$. Clearly, ${P_{\mathrm{P}}}({{\bf{r}}_0},t)$ oscillates between positive and negative values, while $P^{{\mathrm{abs}}}({{\bf{r}}_0},t)$ and $\alpha {S_z}({{\bf{r}}_0},t)$ are always positive. In this single interface setup, the transmitted wave in LT-GaAs is monochromatic and simply propagates in the $z$ direction without reflection. Hence, ${S_z}({{\bf{r}}_0},t)$ is always positive, meaning that the power flux of the transmitted wave is always pointing to the positive $z$ direction (note that ${S_x} = {S_y} = 0$). ${P_{\mathrm{P}}}({{\bf{r}}_0},t)$ shows negative values because of the reactive power. Nevertheless, the time average (over more than one time period) of all three methods are expected to be the same since the time average of the reactive power included in ${P_{\mathrm{P}}}({\bf{r}},t)$ is zero. Indeed, after reaching the steady state, time-averaged ${P^{{\mathrm{abs}}}}({{\bf{r}}_0},t)$, $\alpha |{\bf{S}}({\bf{r}_0},t)|$, and ${P_{\mathrm{P}}}({{\bf{r}}_0},t)$ yields $7.67 \times {10^2}$ ${\mathrm{W/}}{{\mathrm{m}}^{\mathrm{3}}}$, $7.60 \times {10^2}$ ${\mathrm{W/}}{{\mathrm{m}}^{\mathrm{3}}}$, and $7.67 \times {10^2}\,{\mathrm{W/}}{{\mathrm{m}}^{\mathrm{3}}}$, respectively. This validates that both ${P^{{\mathrm{abs}}}}({\bf{r}},t)$ and $\alpha |{\bf{S}}({\bf{r}},t)|$ correctly represent all dissipated power included in ${P_{\mathrm{P}}}({\bf{r}},t)$ in this single interface setup.

To see the difference between the proposed method and the Poynting vector-based method when dealing with non-monochromatic waves, consider a plane wave source with a time signal $sin(2\pi\nu t)+A'sin(2\pi \nu' t)$, and let $\nu$ remain the same as above, $A'=0.2$ and $\nu'=0.01\nu$. The absorption coefficient at $\nu'$ is $\alpha'= 6.06 \times {10^3}\,{{\mathrm{m}}^{ - 1}}$. Since $\alpha'$ at $\nu'$ is two orders of magnitude smaller than $\alpha$ at $\nu$, and the source amplitude at $\nu'$ is $5$ times smaller than that at $\nu$ ($A'=0.2$), the absorbed power density corresponding to $\nu'$ is ignorable and  ${P^{{\mathrm{abs}}}}({{\bf{r}}},t)$ is expected to be almost the same as in the previous example.

Fig.~\ref{abs0} (c) shows the recorded ${P^{{\mathrm{abs}}}}({{\bf{r}}_0},t)$, $\alpha {S_z}({{\bf{r}}_0},t)$ and ${P_{\mathrm{P}}}({{\bf{r}}_0},t)$ at ${{\bf{r}}_0} = (0,0,0)$. As expected, ${P^{{\mathrm{abs}}}}({{\bf{r}}_0},t)$ remains almost the same as in the previous example. However, the time-dependent Poynting vector ${S_z}({{\bf{r}}_0},t)$ shows the beat pattern contributed by the two frequencies $\nu$ and $\nu'$. The time-averaged ${P^{{\mathrm{abs}}}}({{\bf{r}}_0},t)$, $\alpha |{\bf{S}}({\bf{r}_0},t)|$, and ${P_{\mathrm{P}}}({{\bf{r}}_0},t)$ (averaged over the time range $1/\nu'$) are $7.71 \times {10^2}$ ${\mathrm{W/}}{{\mathrm{m}}^{\mathrm{3}}}$, $7.96 \times {10^2}$ ${\mathrm{W/}}{{\mathrm{m}}^{\mathrm{3}}}$, and $7.71 \times {10^2}$ ${\mathrm{W/}}{{\mathrm{m}}^{\mathrm{3}}}$, respectively. This validates that the proposed method can distinguish the power absorbed by the optoelectronic material from that of the low frequency component. The time-dependent Poynting vector contains the power flux of all frequency components and the corresponding absorbed power density is less accurate.

\subsection{Optical Absorption in a Slab}
\label{validation1}

\begin{figure}[ht!]
	\centering
    \begin{tabular}[b]{c}
    \includegraphics[width=0.6\columnwidth]{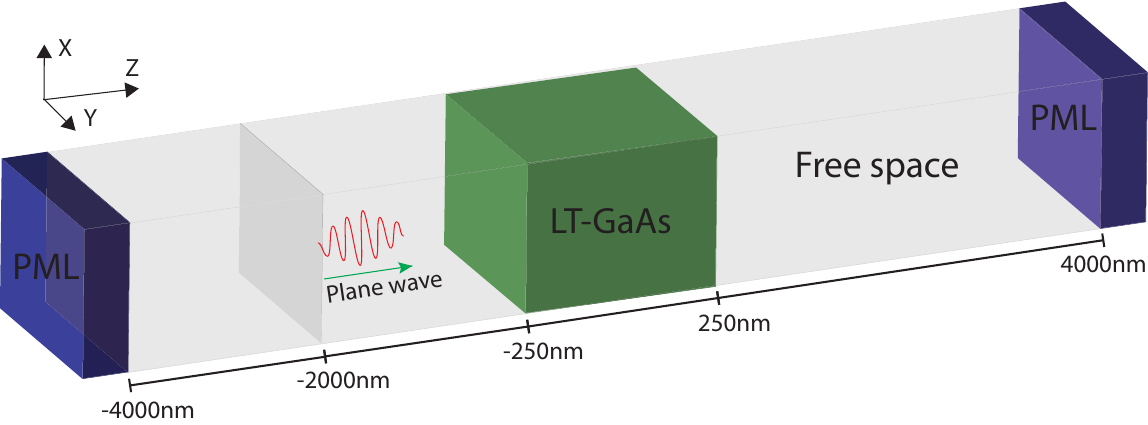}\\
    \footnotesize (a)
    \end{tabular}
    
    \begin{tabular}[b]{c}
    \includegraphics[width=0.45\columnwidth]{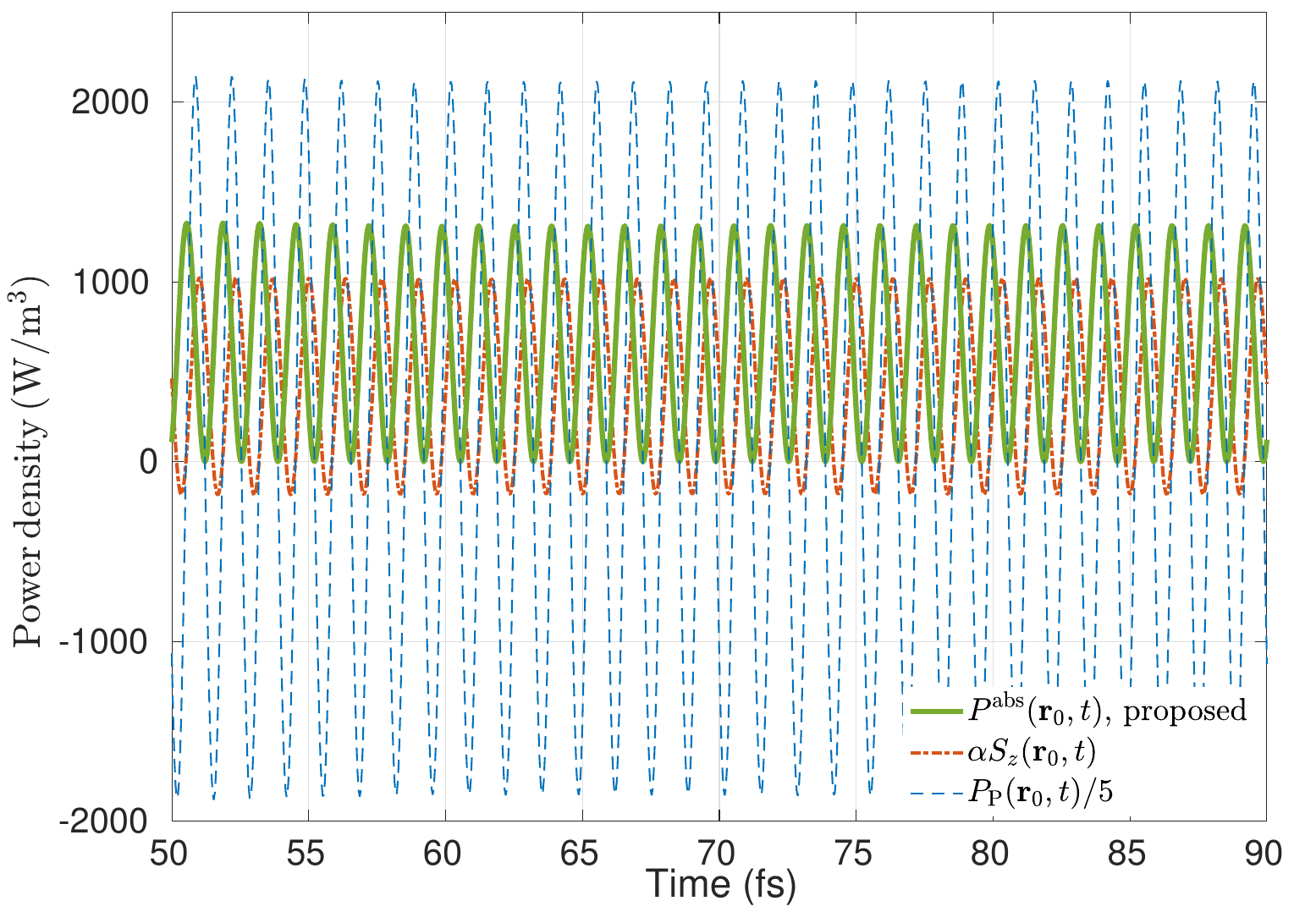}\\
    \footnotesize (b)
    \end{tabular}

    \begin{tabular}[b]{c}
	\includegraphics[width=0.45\columnwidth]{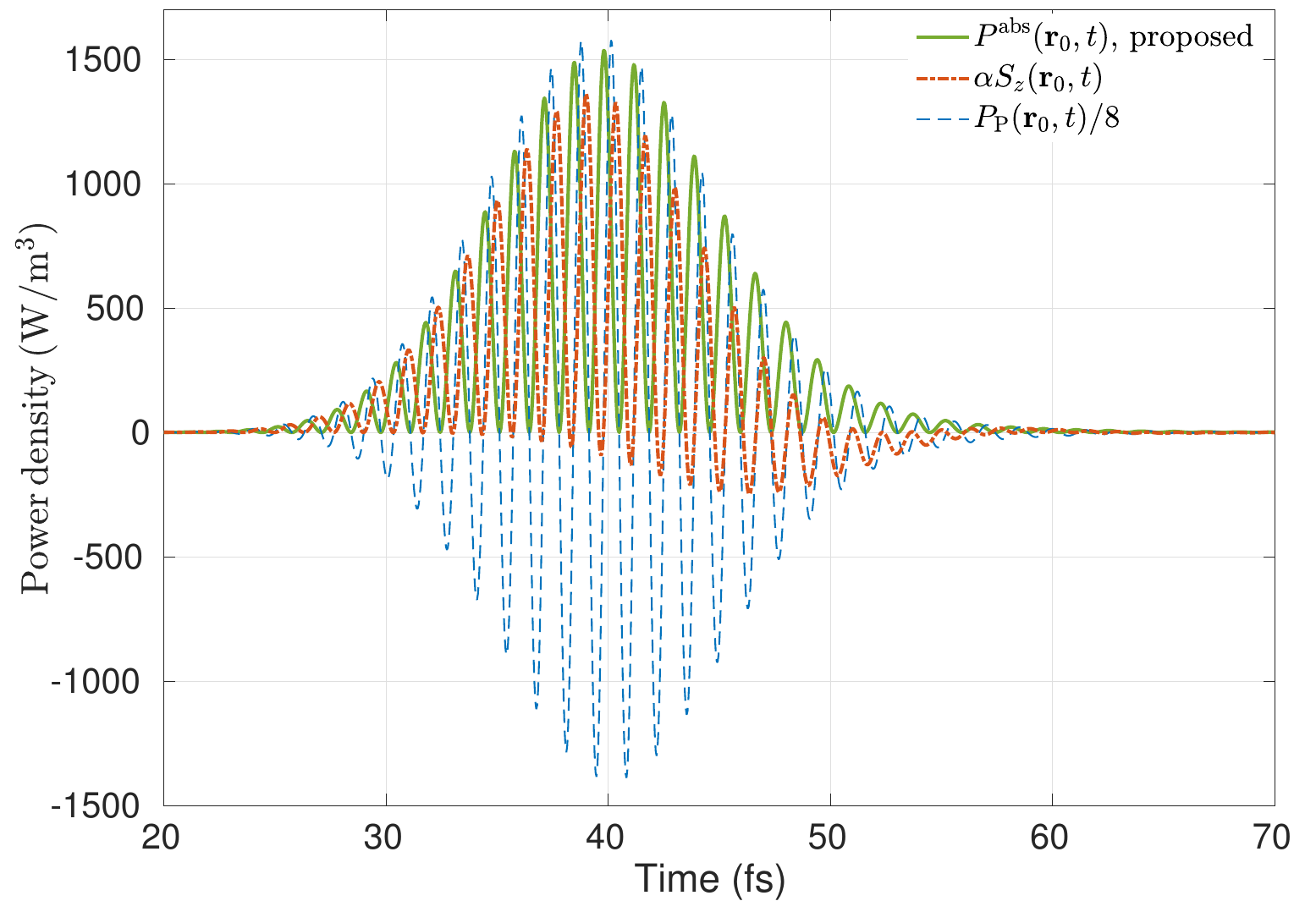}\\
     \footnotesize (c)
     \end{tabular}
    \begin{tabular}[b]{c}
    \includegraphics[width=0.45\columnwidth]{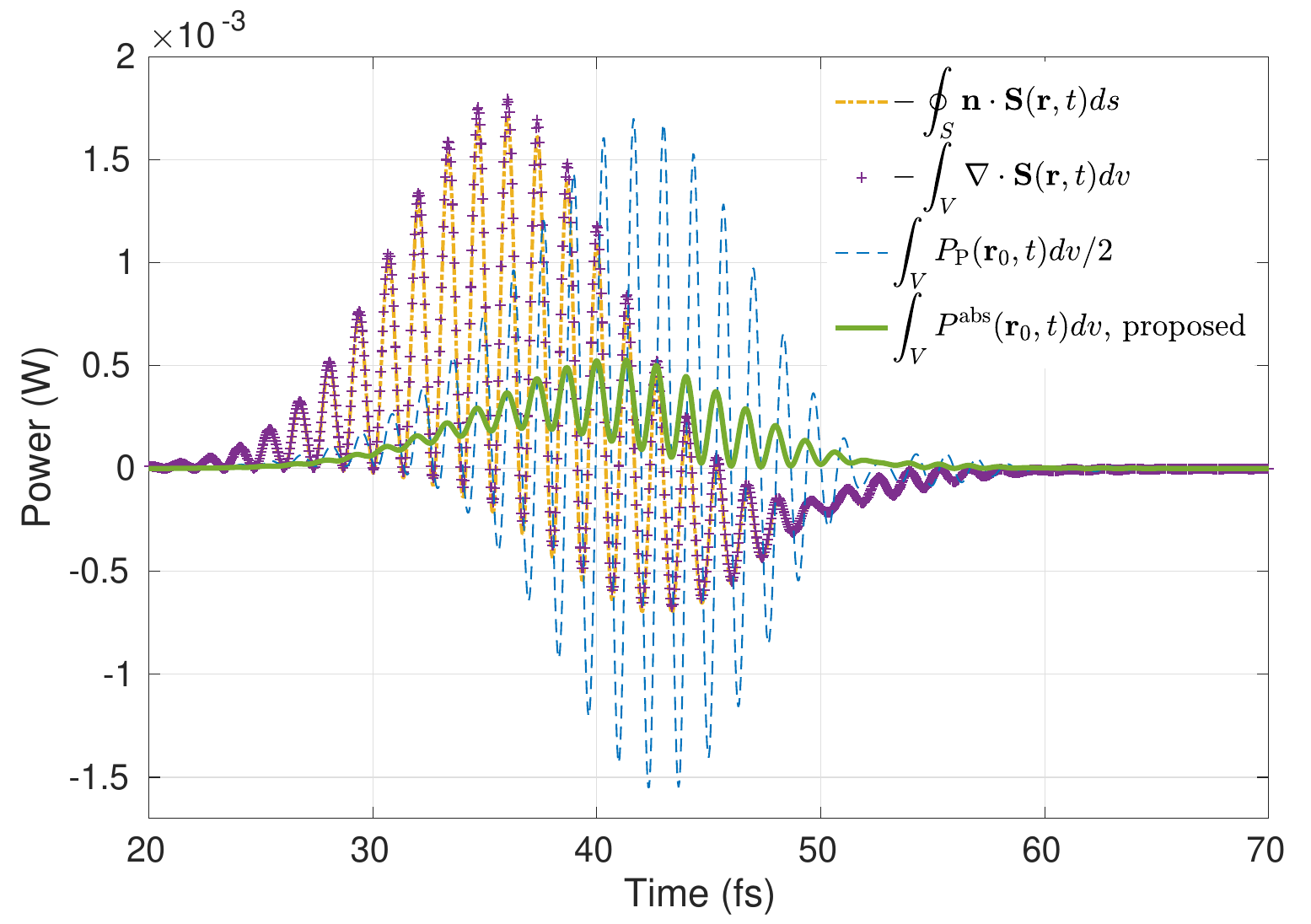}\\
    \footnotesize (d)
    \end{tabular}
	\caption{Simulation with a $500~\mathrm{nm}$-thick LT-GaAs layer. (a) Geometry configuration. ${P^{{\mathrm{abs}}}}({{\bf{r}}_0},t)$ [calculated using~\eqref{Pabs1}], $\alpha {S_z}({{\bf{r}}_0},t)$ and ${P_{\mathrm{P}}}({{\bf{r}}_0},t)$ at ${{\bf{r}}_0} = (0,0,0)$ (b) under a monochromatic excitation and (c) under a Gaussian pulse excitation. (d) Instantaneous absorbed power in the LT-GaAs layer under a Gaussian pulse excitation. Note that ${P_{\mathrm{P}}}({{\bf{r}}_0},t)$ is scaled for a better demonstration.}
	\label{abs}
\end{figure}

To further validate the proposed method and see the limitations of the Poynting vector-based method, consider a $500\,{\mathrm{nm}}$-thick LT-GaAs layer resides in free space. The material parameters are the same as before. The simulation setup is depicted in Fig.~\ref{abs}(a). A monochromatic $x$-polarized plane wave with frequency $\nu = 375\,{\mathrm{THz}}$ is normally incident on the LT-GaAs layer. 

Fig.~\ref{abs} (b) shows ${P^{{\mathrm{abs}}}}({{\bf{r}}_0},t)$ calculated using the proposed method, $\alpha {S_z}({{\bf{r}}_0},t)$, and ${P_{\mathrm{P}}}({{\bf{r}}_0},t)$ at ${{\bf{r}}_0}\! =\! (0,0,0)$. The figure shows that $P^{{\mathrm{abs}}}({{\bf{r}}_0},t)$ calculated using~\eqref{Pabs1} is always positive, while $\alpha {S_z}({{\bf{r}}_0},t)$ and ${P_{\mathrm{P}}}({{\bf{r}}_0},t)$ are oscillating between positive and negative values. Here, the negative value of ${S_z}({{\bf{r}}_0},t)$ means that the instantaneous power flux is pointing in the $- z$ direction due to the reflection on the interface at $z = 250\,{\mathrm{nm}}$.
Like before, the oscillations in ${P_{\mathrm{P}}}({{\bf{r}}_0},t)$ are caused by the reactive power. 
After reaching the steady state, time-averaged ${P^{{\mathrm{abs}}}}({{\bf{r}}_0},t)$, $\alpha |{\bf{S}}({\bf{r}_0},t)|$, and ${P_{\mathrm{P}}}({{\bf{r}}_0},t)$ yields $6.57 \times {10^2}$ ${\mathrm{W/}}{{\mathrm{m}}^{\mathrm{3}}}$, $4.80 \times {10^2}$ ${\mathrm{W/}}{{\mathrm{m}}^{\mathrm{3}}}$, and $6.57 \times {10^2}\,{\mathrm{W/}}{{\mathrm{m}}^{\mathrm{3}}}$, respectively. This again validates that ${P^{{\mathrm{abs}}}}({\bf{r}},t)$ correctly represents all dissipated power included in ${P_{\mathrm{P}}}({\bf{r}},t)$. It also indicates that using the time-domain Poynting vector to calculate the instantaneous absorbed power density is not accurate in general scattering problems where the direction of the local power flux is time-dependent.

Next, the same test is performed with a wideband pulsed source. A Gaussian pulse signal,
\begin{equation*}
f(t) = {e^{ - {{(t - {t_0})}^2}/{\tau ^2}}}sin(2\pi \nu t)
\end{equation*}
where $\nu \! =\! 375\,{\mathrm{THz}}$, $\tau \! =\! 10\,{\mathrm{fs}}$, and ${t_0} \! =\! 3\tau $, is used. Note that in real applications, e.g., see the example in the next subsection, $\tau$ is usually much larger (at the scale of $1\,\mathrm{ps}$). Here, a wideband signal is chosen for better demonstration and also to show that the proposed method can deal with wideband excitation cases. Fig.~\ref{abs} (c) shows ${P^{{\mathrm{abs}}}}({{\bf{r}}_0},t)$, $\alpha {S_z}({{\bf{r}}_0},t)$, and ${P_{\mathrm{P}}}({{\bf{r}}_0},t)$. Again, ${P^{{\mathrm{abs}}}}({{\bf{r}}_0},t)$ remains positive during the simulation, while $\alpha S_z({{\bf{r}}_0},t)$, and ${P_{\mathrm{P}}}({{\bf{r}}_0},t)$ take on negative values. The accumulated power densities computed by integrating ${P^{{\mathrm{abs}}}}({{\bf{r}}_0},t)$, $\alpha |{\bf{S}}({{\bf{r}}_0},t)|$, and ${P_{\mathrm{P}}}({{\bf{r}}_0},t)$ are $2.45\,{\mathrm{J/}}{{\mathrm{m}}^{\mathrm{3}}}$, $2.01\,{\mathrm{J/}}{{\mathrm{m}}^{\mathrm{3}}}$, and $2.45\,{\mathrm{J/}}{{\mathrm{m}}^{\mathrm{3}}}$, respectively, showing that ${P^{{\mathrm{abs}}}}({{\bf{r}}_0},t)$ of the proposed method correctly represents the (local) absorbed power density while $\alpha |{\bf{S}}({{\bf{r}}_0},t)|$ is less accurate.

Fig.~\ref{abs} (d) shows the total absorbed power in the LT-GaAs layer, where $V$ and $S$ represents its volume and surface, respectively, and ${\bf{\hat n}}({\bf{r}})$ is the outward pointing unit normal vector on $S$. Using the Poynting theorem, one can see that both $ - \int_V {\nabla \cdot {\bf{S}}({\bf{r}},t)dv} $ and $ - \int_S {{\bf{\hat n}}({\bf{r}}) \cdot {\bf{S}}({\bf{r}},t)ds}$ yield the instantaneous net power entering volume $V$, and $\int_V {{P_{\mathrm{P}}}({\bf{r}},t)dv} $ corresponds to the mechanic work done during the polarization process. The values of all these three expressions oscillate due to the reactive power. Their negative ``tails'' at the late time (after around $40\,{\mathrm{fs}}$) mean that the pulse energy gradually leaves the LT-GaAs layer. More importantly, $\int_V {{P^{{\mathrm{abs}}}}({\bf{r}},t)dv} $ is always positive, and the total absorbed energy calculated by integrating these four expressions over time is the same and equal to $4.80 \times {10^{-18}}\,{\mathrm{J}}$. This example shows that the proposed method works well for a wideband excitation.

\subsection{Carrier Generation in Photoconductive Devices}
\begin{figure}[ht!]
	\centerline{\includegraphics[width=0.6\columnwidth]{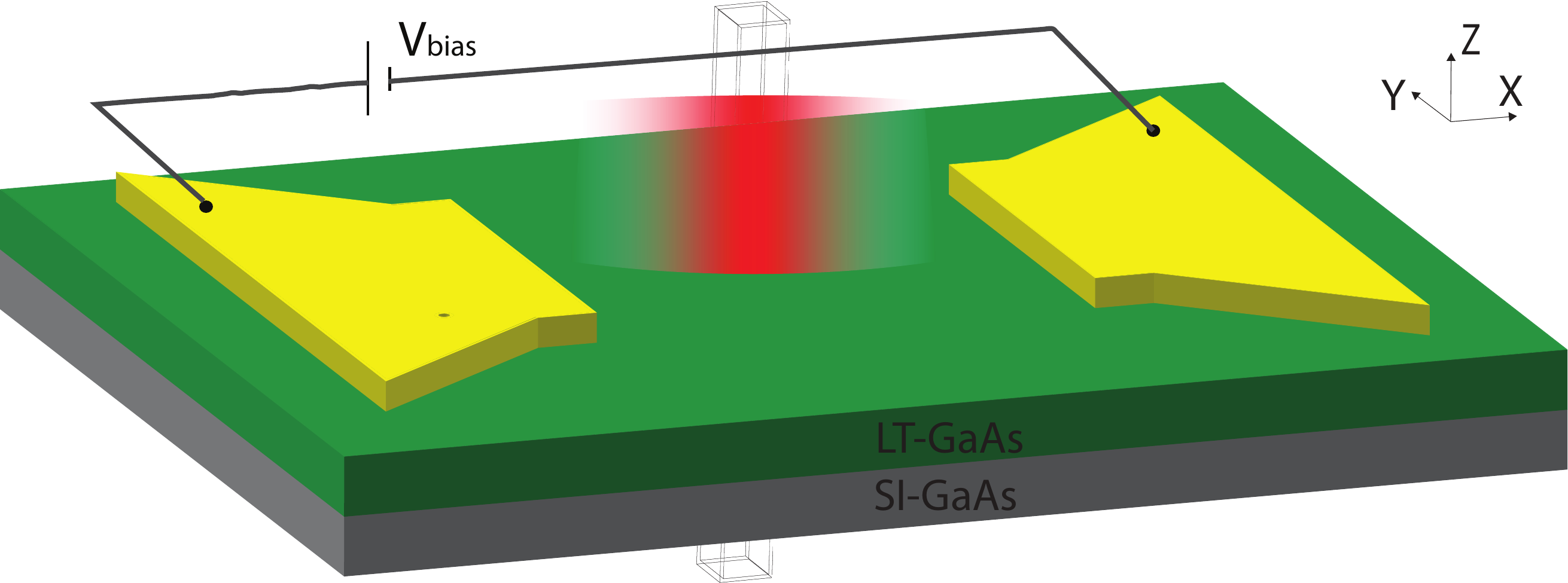}}
	\caption{Schematic of the photoconductive device. The gray box is the domain of the unit-cell model. Note that the size of the THz antenna attached to the electrodes shown in the figure is much larger than the gap distance between the electrodes.}
	\label{PCD}
\end{figure}

Next, the proposed method is used in the simulation of the photoconductive device shown in Fig.~\ref{PCD}. Both the LT-GaAs photoconductive layer and the semi-insulating GaAs (SI-GaAs) substrate are $500\,{\mathrm{nm}}$ thick, and the interface between these two layers is located on the $xy$ plane. A bias voltage ${V_{{\mathrm{bias}}}}$ is applied to the electrodes. The distance between the electrodes (along the $x$ direction) is $5\,\mu {\mathrm{m}}$. For LT-GaAs, the permittivity and permeability are the same as those used in the previous example and the semiconductor material parameters are the same as those in~\cite{Chen2019multiphysics}. The relative permittivity of SI-GaAs is $13.26$. 

To carry out the discretization in space, the unit-cell-based discontinuous Galerkin method developed in~\cite{Chen2020efficient} is used. First, the steady state of the semiconductor device biased with ${V_{{\mathrm{bias}}}}$ is simulated by solving the coupled system of the Poisson equation and the stationary drift-diffusion equations~\cite{Chen2020steadystate}. For the Poisson equation, a potential-drop boundary condition (to mimic the effect of the bias voltage), the periodic boundary conditions, and the homogeneous Neumann boundary condition are used along the $x$, $y$, and $z$ directions, respectively. For the stationary drift-diffusion equation, the periodic boundary conditions are used along the $x$ and $y$ directions, and the homogeneous Robin boundary condition is enforced on the surfaces of the LT-GaAs layer along the $z$ direction~\cite{Chen2020float, Chen2020hybridizable}. The steady-state electric field [denoted as ${{\bf{E}}^s}({\bf{r}})$ in~\eqref{t_J0}], which is obtained by solving this coupled system of the Poisson equation and the stationary drift-diffusion equations, and the corresponding field-dependent mobility [denoted as ${\mu _c}({\bf{r}})$, $c \in \{ e,h\} $, in~\eqref{t_J0}] are used as inputs for the time-domain simulation. This simulation solves the coupled system of the time-domain Maxwell equations~\eqref{t_E0}-\eqref{t_H0} and the time-domain drift-diffusion equations~\eqref{t_N0}-\eqref{t_J0} and uses~\eqref{generation} to compute $G({\bf{E}},{\bf{H}})$~\cite{Chen2019multiphysics, Chen2020efficient}. For both the Maxwell and the drift-diffusion equations, the periodic boundary conditions are used along the $x$ and $y$ directions. Along the $z$ direction, perfectly matched layers~\cite{Chen2020pml, Gedney2009} are used for the Maxwell equations and the homogeneous Robin boundary condition is used for the drift-diffusion equations. Two types of time-domain simulations are carried out: One that solves~\eqref{t_E0}-\eqref{generation} and uses~\eqref{Pabs1} to compute ${P^{{\mathrm{abs}}}}({\bf{r}},t)$ (with the time integration described in Section~\ref{formu}) and another one that solves~\eqref{t_E0}-\eqref{generation} but uses~\eqref{PabsS1} to compute ${P^{{\mathrm{abs}}}}({\bf{r}},t)$. 

\begin{figure}[ht!]
	\centering
    \begin{tabular}[b]{c}
	\includegraphics[width=0.55\columnwidth]{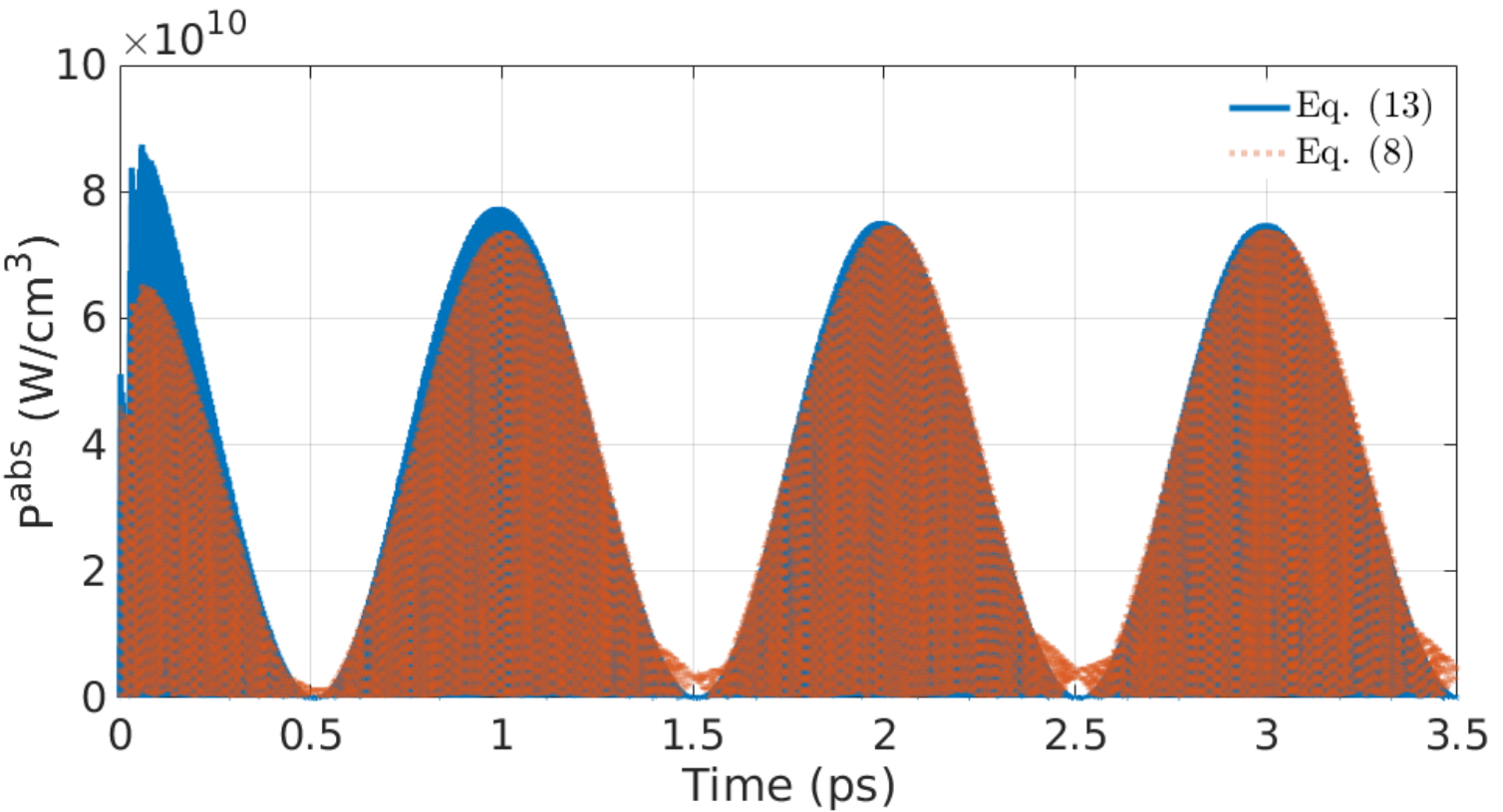}\\
    \footnotesize (a)
    \end{tabular}
    
    \begin{tabular}[b]{c}
    \includegraphics[width=0.55\columnwidth]{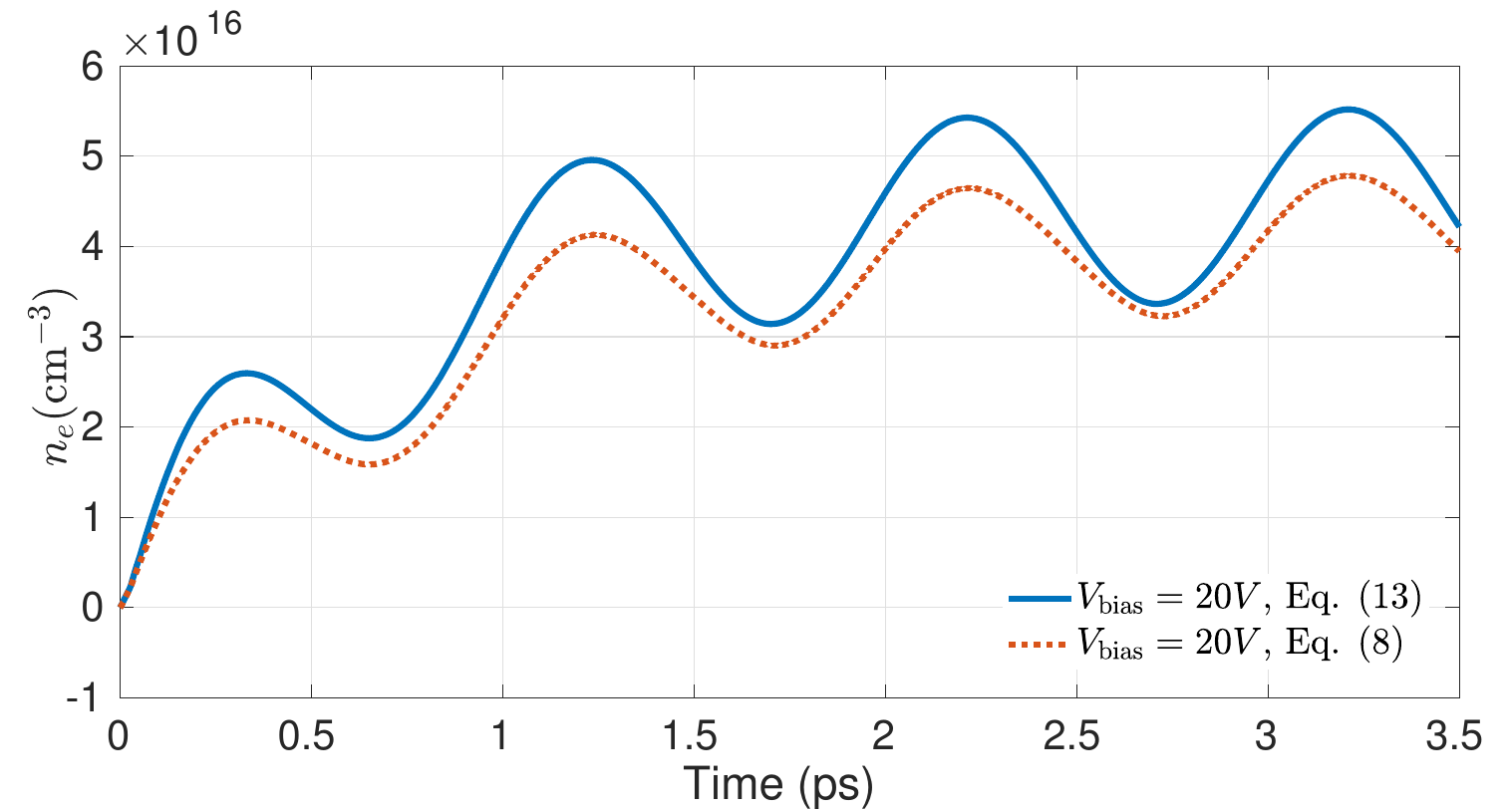}\\
    \footnotesize (b)
    \end{tabular}
	\caption{(a) ${P^{{\mathrm{abs}}}}({{\bf{r}}_1},t)$ and (b) ${n_e}({{\bf{r}}_1},t)$ computed by the two time-domain simulations that use~\eqref{PabsS1} and~\eqref{Pabs1} at ${{\bf{r}}_1} = (0,0,480)\,{\mathrm{nm}}$ for ${V_{{\mathrm{bias}}}} = 20\,{\mathrm{V}}$.}
	\label{LowV}
\end{figure}

The photoconductive device is operated in the continuous-wave mode~\cite{Lepeshov2017review} and excited from top by two continuous-wave $x$-polarized lasers operating at $374.5\,{\mathrm{THz}}$ and $375.5\,{\mathrm{THz}}$. Under this excitation, the photocarrier magnitude varies with a frequency of $1\,{\mathrm{THz}}$, leading to the generation of THz electromagnetic fields~\cite{Lepeshov2017review}. For low values of ${V_{{\mathrm{bias}}}}$ and low levels of laser power,~\eqref{PabsA}-\eqref{PabsS1}, which are used to compute $G({\bf{E}},{\bf{H}})$, have been validated and are found to agree with each other~\cite{Chen2019multiphysics, Moreno2014, Burford2016, Khorshidi2016}. For the first set of simulations, ${V_{{\mathrm{bias}}}}\! =\! 20\,{\mathrm{V}}$ and the laser power flux density ($S_{\mathrm{pump}}$) is $3.32 \times 10^{9}\; \mathrm{mW/cm^2}$ (both of these values are relatively low). Fig.~\ref{LowV} (a) shows ${P^{{\mathrm{abs}}}}({{\bf{r}}_1},t)$ computed by the two simulations [one that uses~\eqref{Pabs1} and the other one that uses~\eqref{PabsS1}], where ${{\bf{r}}_1}\! =\! (0,0,480)\,{\mathrm{nm}}$. Results are similar, however, ${P^{{\mathrm{abs}}}}({{\bf{r}}_1},t)$ computed using~\eqref{PabsS1} is less smooth (see the data near $t = 1.5$, $2.5$, and $3.5\,{\mathrm{ps}}$) because of using the time-dependent Poynting vector. Fig.~\ref{LowV} (b) shows that the electron density ${n_e}({{\bf{r}}_1},t)$ computed by the same two simulations behave in a similar way. Since ${n_e}({\bf{r}},t)$ changes exponentially with ${P^{{\mathrm{abs}}}}({\bf{r}},t)$ [via $G({\bf{E}},{\bf{H}})$], the larger relative difference between the results in Fig.~\ref{LowV} (b) as opposed to that in Fig.~\ref{LowV} (a) is expected. 

\begin{figure}[ht!]
	\centering
    \begin{tabular}[b]{c}
    \includegraphics[width=0.55\columnwidth]{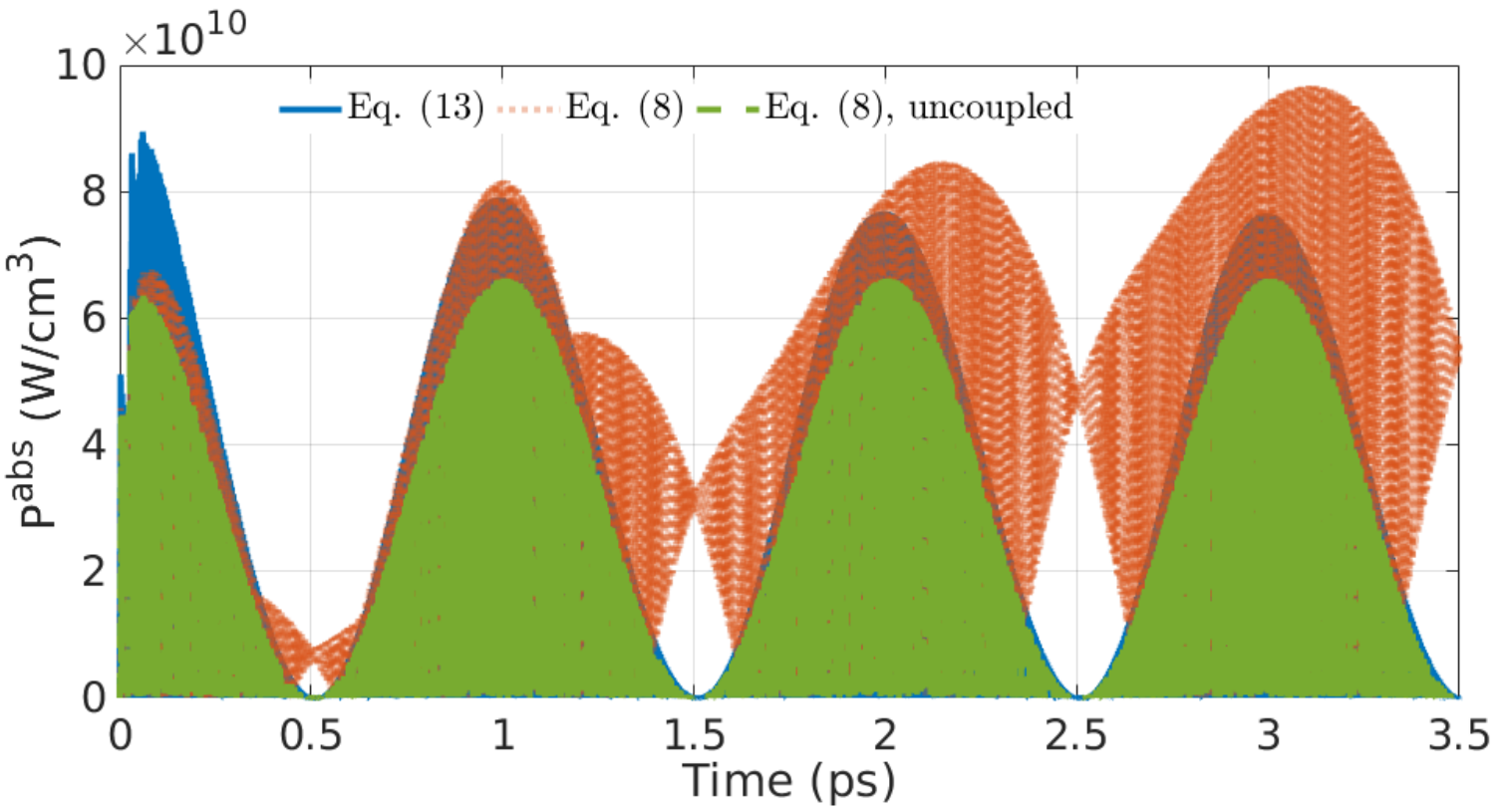}\\
    \footnotesize (a)
    \end{tabular}

    \begin{tabular}[b]{c}
    \includegraphics[width=0.55\columnwidth]{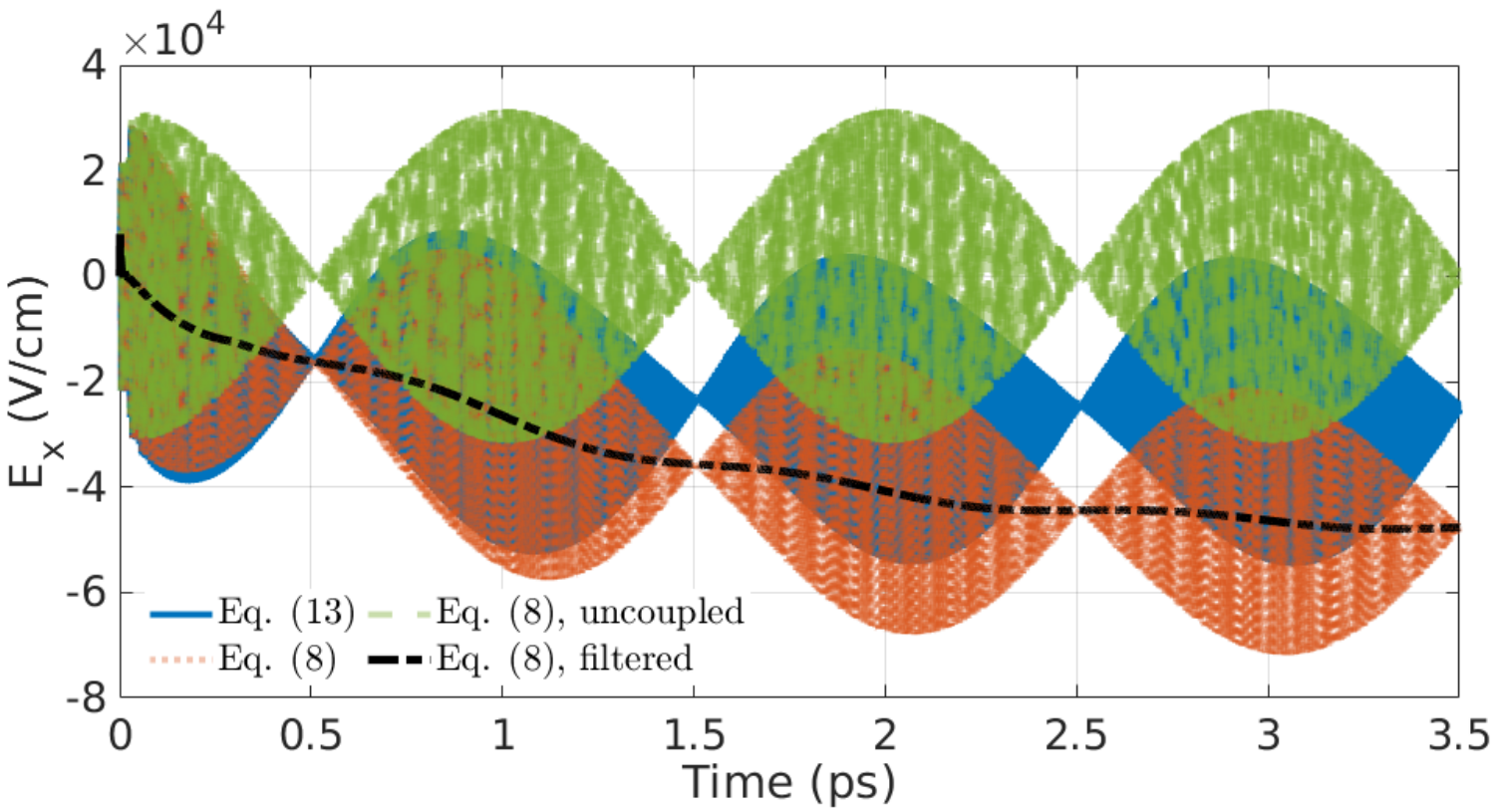}\\
    \footnotesize (b)
    \end{tabular}

    \begin{tabular}[b]{c}
    \includegraphics[width=0.55\columnwidth]{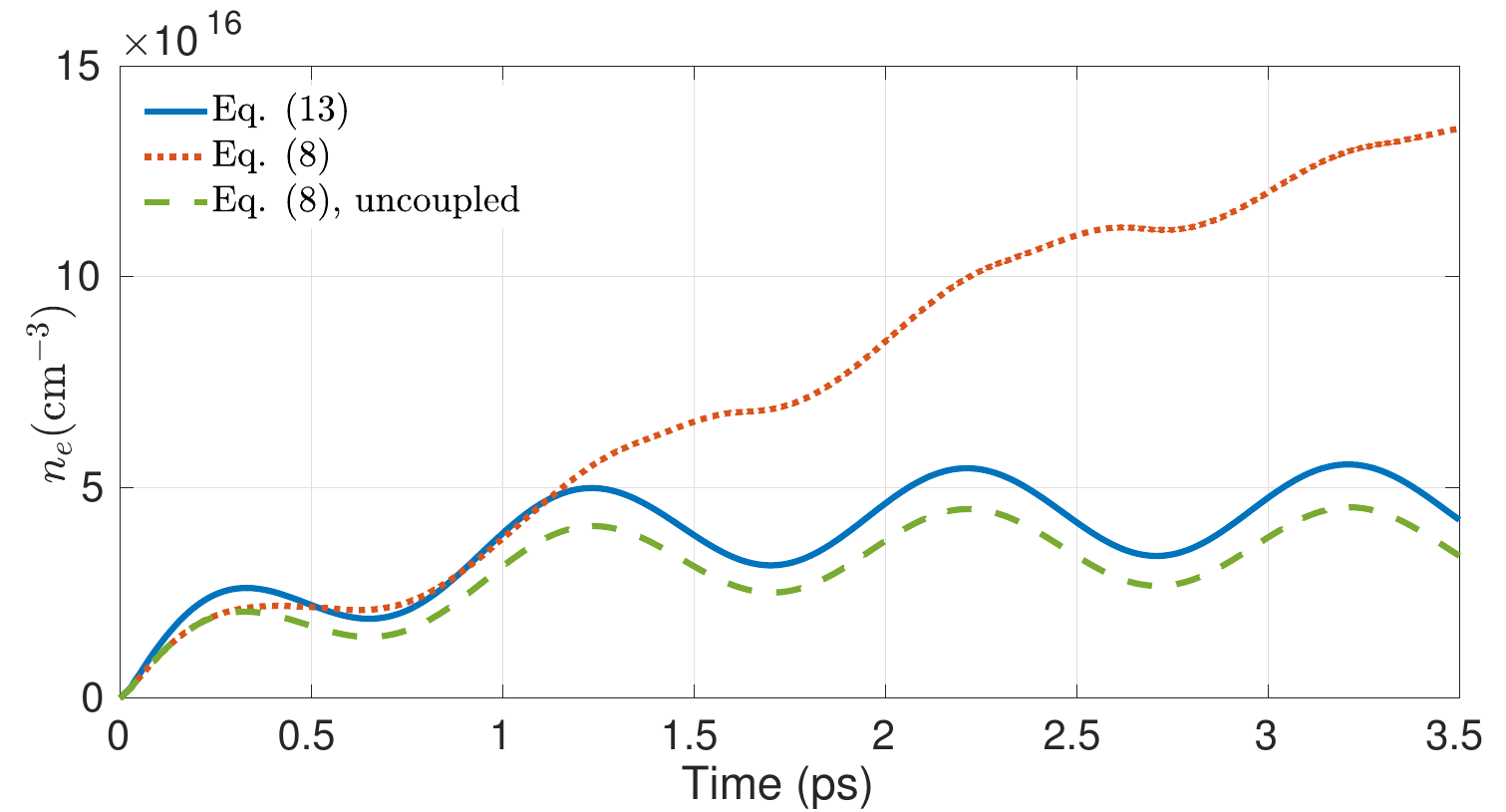}\\
    \footnotesize (c)
    \end{tabular}
    \caption{(a) ${P^{{\mathrm{abs}}}}({{\bf{r}}_1},t)$, (b) $x$-component of ${\bf{E}}({{\bf{r}}_1},t)$, and (c) ${n_e}({{\bf{r}}_1},t)$ computed by the two time-domain simulations that use~\eqref{PabsS1} and~\eqref{Pabs1} and the ``uncoupled'' time-domain simulation that use~\eqref{PabsS1} at ${{\bf{r}}_1} = (0,0,480)\,{\mathrm{nm}}$ for ${V_{{\mathrm{bias}}}} = 40{\mathrm{\;V}}$.}
	\label{HighV}
\end{figure}

The photocurrent density ${{\bf{J}}_{\mathrm{D}}}({\bf{r}},t)$ depends on ${V_{{\mathrm{bias}}}}$ and the power level of the laser. The photoconductive device generates stronger low-frequency electromagnetic fields when excited by a higher-power laser [which results in a higher number of photocarriers and a higher ${{\bf{J}}_{\mathrm{D}}}({\bf{r}},t)$] and/or biased by a higher $V_{\mathrm{bias}}$ [which provides a larger drift force and increases ${{\bf{J}}_{\mathrm{D}}}({\bf{r}},t)$]. Since the Poynting vector $\mathbf{S}(\mathbf{r},t)$ also includes the power due to these low-frequency fields,~\eqref{PabsS1} overestimates $G({\bf{E}},{\bf{H}})$. To demonstrate this problem clearly, the simulations described above are repeated for ${V_{{\mathrm{bias}}}}\! =\! 40\,{\mathrm{V}}$ (the power level of the laser is kept the same). Fig.~\ref{HighV} (a) shows that ${P^{{\mathrm{abs}}}}({\bf{r}},t)$ computed using~\eqref{PabsS1} continues to increase and eventually becomes much larger than the maximally possible absorbed power density estimated from the laser source and the absorption coefficient, which is $2 \alpha S_{\mathrm{pump}} = 6.64\times 10^{10} \;\mathrm{W/cm^3}$, where $2$ comes from the two laser sources. Clearly, this is unphysical because the source power remains unchanged during the simulation. In contrast, ${P^{{\mathrm{abs}}}}({\bf{r}},t)$ computed using~\eqref{Pabs1} behaves as expected: It stays at a stationary level once the field and the carrier interactions on the device reach the steady state. 

Fig.~\ref{HighV} (b) shows ${\bf{E}}({{\bf{r}}_1},t)$ computed in the same simulations. The high-frequency oscillation of ${\bf{E}}({{\bf{r}}_1},t)$ corresponds to the optical component. The black dash-dotted curve is obtained after a low pass filter (averaging over a sliding window of length $0.25\,{\mathrm{ps}}$) is applied to ${\bf{E}}({{\bf{r}}_1},t)$ recorded during the simulation that uses~\eqref{PabsS1}. In the filtered curve, the ripples (oscillates at a time scale of $1\;\mathrm{ps}$) corresponds to the THz component and the amplitude of the ripples indicates its strength. The nonzero average field corresponds to the near-DC components. This curve indicates that, as expected, low-frequency fields, which are generated by ${{\bf{J}}_{\mathrm{D}}}({\bf{r}},t)$, are also included in ${\bf{E}}({{\bf{r}}_1},t)$~\cite{Chen2021screen}. Comparing Fig.~\ref{HighV} (a) and Fig.~\ref{HighV} (b), it is evident that ${P^{{\mathrm{abs}}}}({\bf{r}},t)$ computed using~\eqref{PabsS1} follows the envelope of the corresponding electric field, which contains the THz component and near-DC components. This consequently means that~\eqref{PabsS1} overestimates ${P^{{\mathrm{abs}}}}({\bf{r}},t)$ since, as discussed in Section~\ref{intro}, the low-frequency fields should not be taken into account in the computation of $G({\bf{E}},{\bf{H}})$. In addition, this overestimated $G({\bf{E}},{\bf{H}})$ lead to a stronger ${{\bf{J}}_{\mathrm{D}}}({\bf{r}},t)$, which then generates stronger THz fields, which finally leads to an even higher $G({\bf{E}},{\bf{H}})$. This feedback loop eventually results in an overestimated carrier density, as demonstrated by ${n_e}({{\bf{r}}_1},t)$ shown in Fig.~\ref{HighV} (c).

For comparison, an ``uncoupled'' simulation, which solves~\eqref{t_E0}-\eqref{generation} and uses~\eqref{PabsS1} but ignores ${{\bf{J}}_{\mathrm{D}}}({\bf{r}},t)$ on the right-hand side of~\eqref{t_E0}, is executed. The results obtained by this simulation are tagged with ``uncoupled'' in Fig.~\ref{HighV}. Since no low-frequency fields are generated when ${{\bf{J}}_{\mathrm{D}}}({\bf{r}},t)$ is removed from the right-hand side of~\eqref{t_E0}, ${P^{{\mathrm{abs}}}}({\bf{r}},t)$ calculated using~\eqref{PabsS1} in this simulation remains stable. This again verifies that ${P^{{\mathrm{abs}}}}({\bf{r}},t)$ and ${n_e}({{\bf{r}}},t)$ go unphysically high in the coupled simulation using~\eqref{PabsS1} is because of that ${\bf{S}}({\bf{r}},t)$ includes the power flux density of the low-frequency fields.

In contrast, in Fig.~\ref{HighV}, ${P^{{\mathrm{abs}}}}({\bf{r}_1},t)$, ${\bf{E}}({{\bf{r}}_1},t)$, and ${n_e}({{\bf{r}}_1},t)$ calculated using the proposed method that solves~\eqref{t_E0}-\eqref{generation} and uses~\eqref{Pabs1}, behave as expected. Since~\eqref{Pabs1} naturally models the absorptance of LT-GaAs as high at optical frequencies and as negligible at THz frequencies (imaginary part of the permittivity being zero), the incorrect behavior of ${P^{{\mathrm{abs}}}}({\bf{r}},t)$ and ${n_e}({{\bf{r}}},t)$, which is present in the coupled simulation that uses~\eqref{PabsS1}, is not observed in this simulation. Meanwhile, the low-frequency fields generated by ${{\bf{J}}_{\mathrm{D}}}({\bf{r}},t)$ are accounted for [Fig.~\ref{HighV} (b)]. This permits numerical analysis of the low-frequency radiation-field screening effect observed experimentally in the response of photoconductive devices~\cite{Chen2020efficient, Darrow1992, Benicewicz1994, Kim2006, Loata2007, Chou2013, Chen2021screen}. Furthermore, in Fig.~\ref{HighV} (c), ${n_e}({{\bf{r}}},t)$ computed in the simulation that uses~\eqref{Pabs1} is slightly higher than that computed in the uncoupled simulation that uses~\eqref{PabsS1}. As has been shown in the examples in Sections~\ref{validation} and~\ref{validation1}, ${P^{{\mathrm{abs}}}}({\bf{r}},t)$ calculated using~\eqref{Pabs1} is more accurate than that calculated using~\eqref{PabsS1}. This suggests that ${n_e}({{\bf{r}}_1},t)$ computed in the simulation that uses~\eqref{Pabs1} is more accurate than that computed in the uncoupled simulation that uses~\eqref{PabsS1}.

\begin{figure}[ht!]
	\centering
    \begin{tabular}[b]{c}
    \includegraphics[width=0.35\columnwidth]{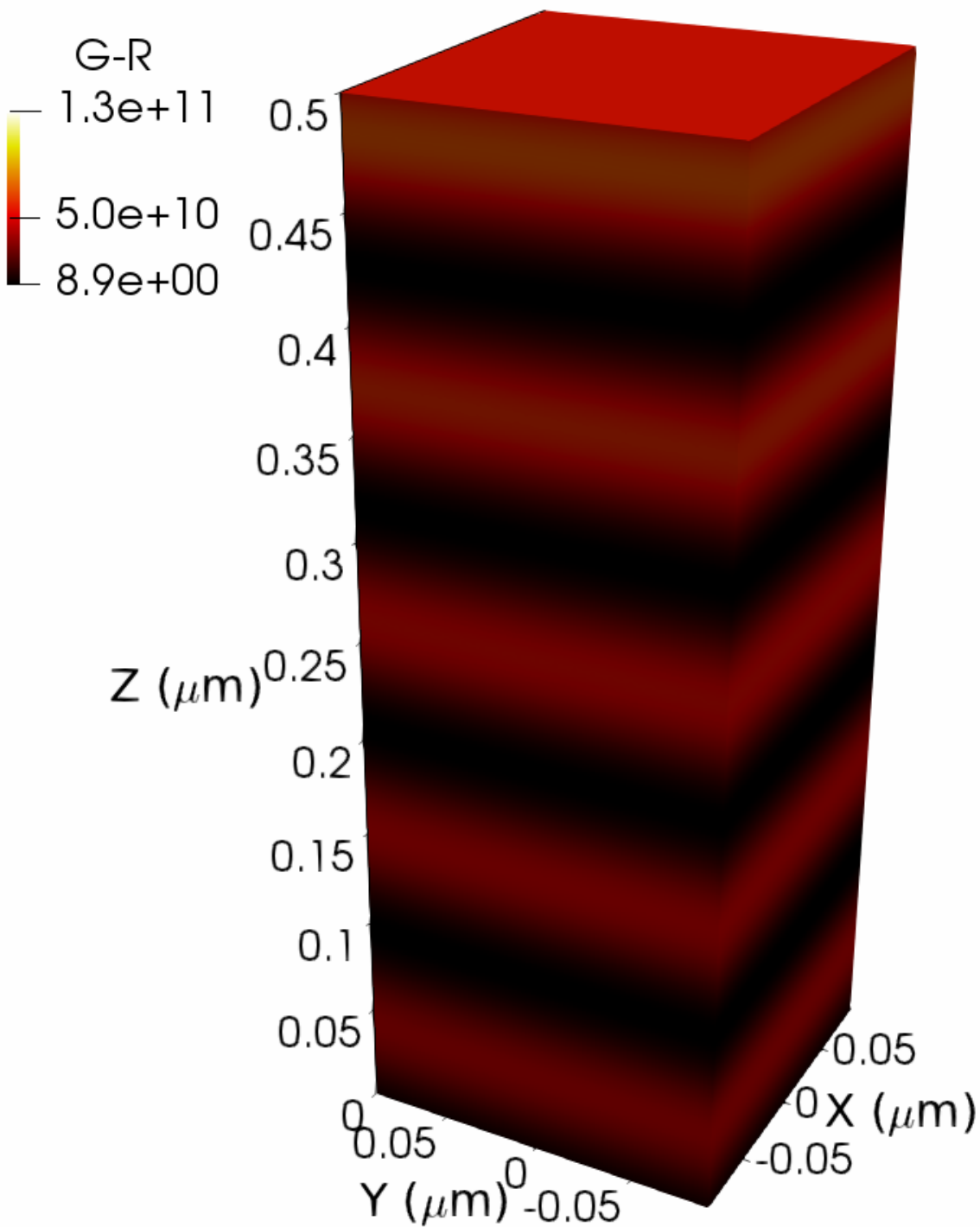}\\
    \footnotesize (a)
    \vspace{2cm}
    \end{tabular}
    \begin{tabular}[b]{c}
    \includegraphics[width=0.35\columnwidth]{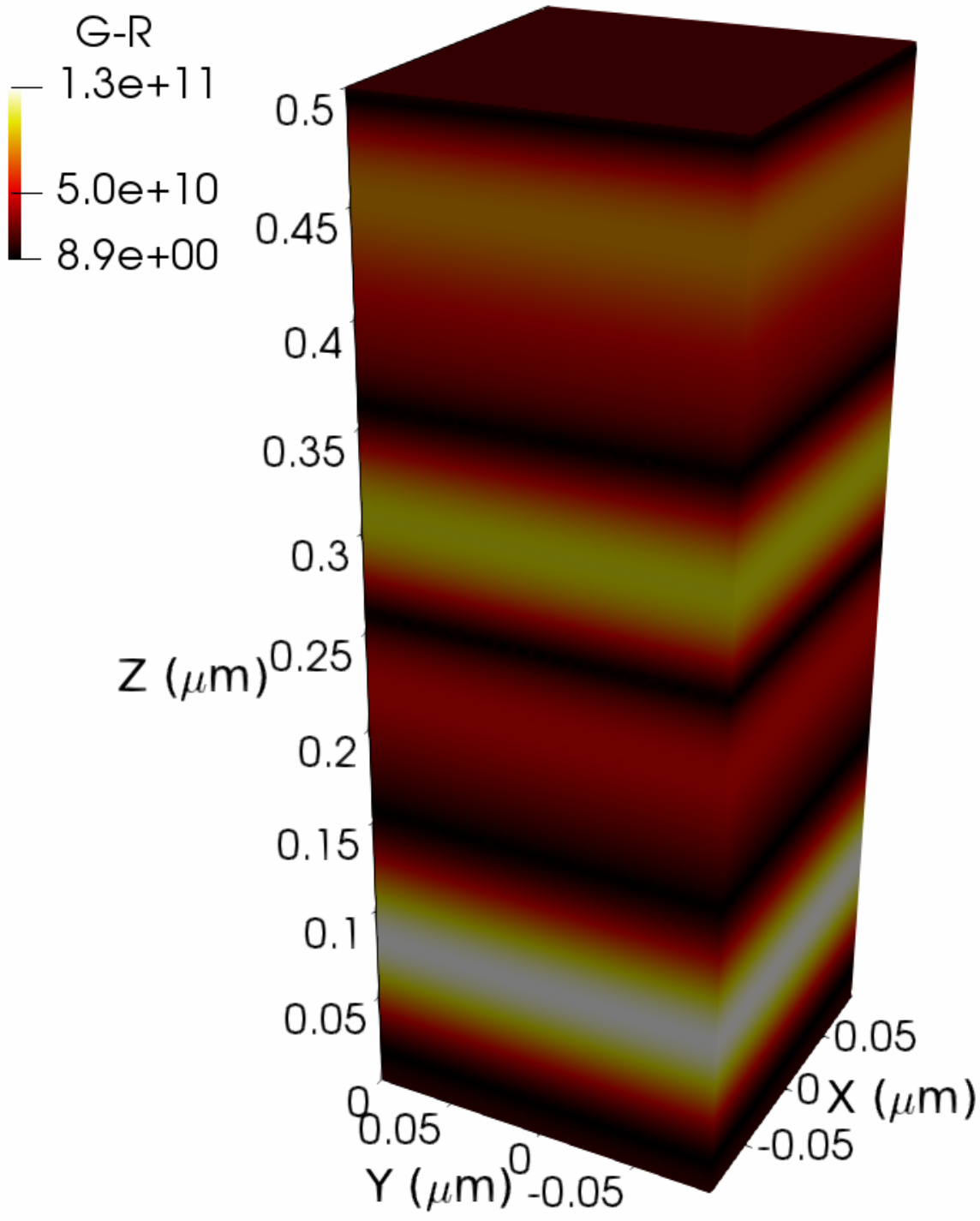}\\
    \footnotesize (b)
    \vspace{2cm}
    \end{tabular}
    
    \begin{tabular}[b]{c}
    \includegraphics[width=0.35\columnwidth]{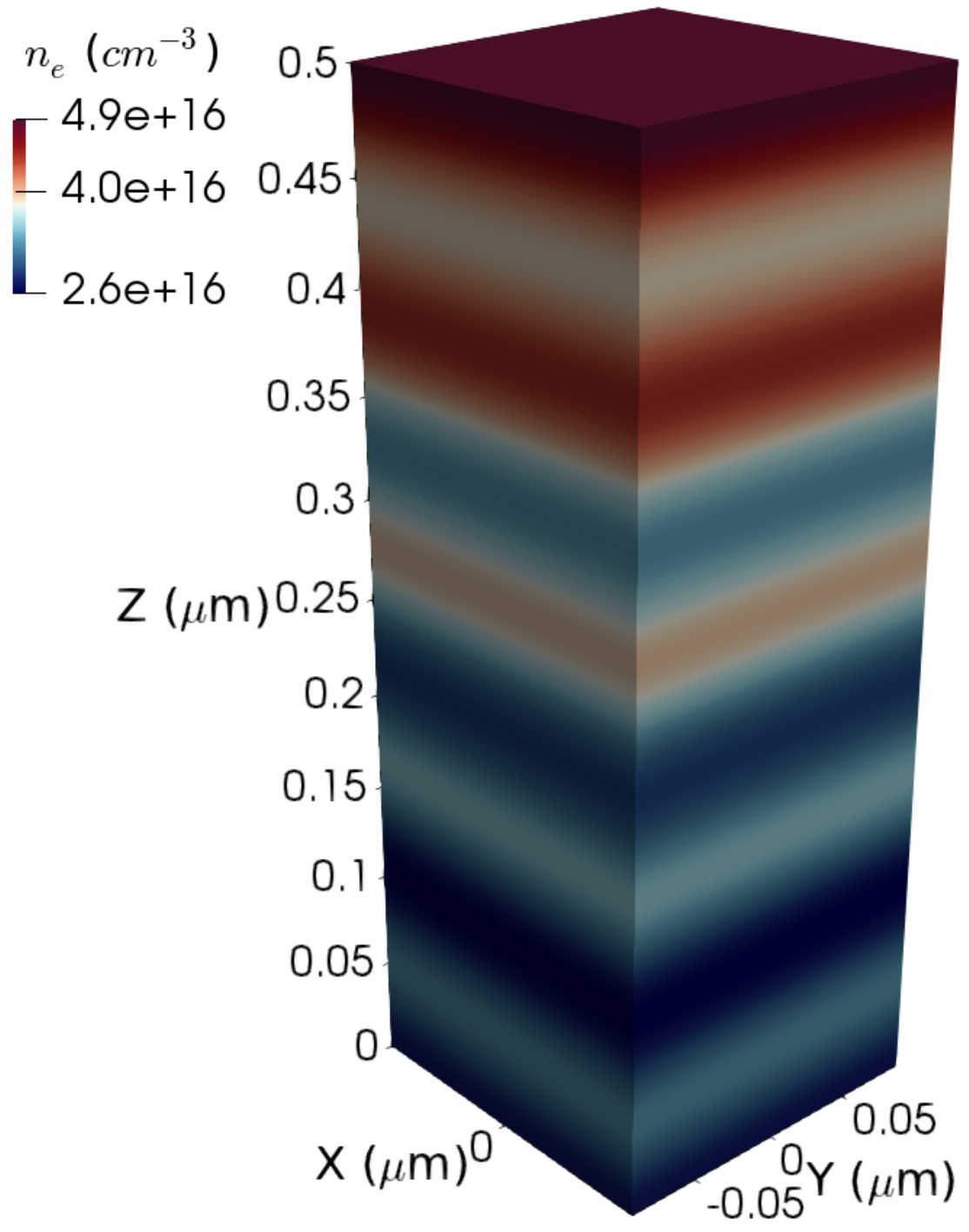}\\
    \footnotesize (c)
    \end{tabular}
    \begin{tabular}[b]{c}
    \includegraphics[width=0.35\columnwidth]{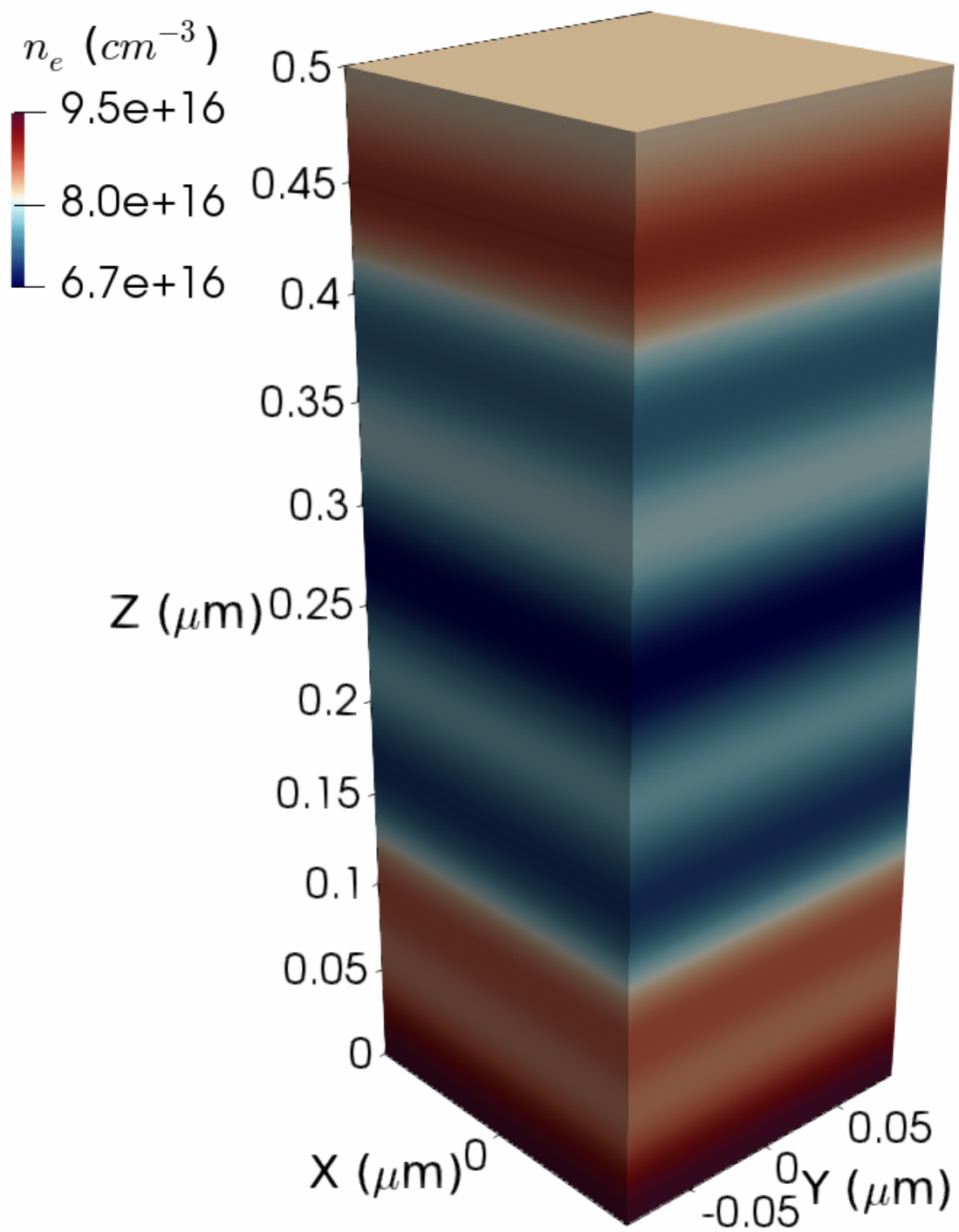}\\
    \footnotesize (d)
    \end{tabular}
	\caption{(a) $[G({\bf{E}},{\bf{H}}) - R({n_e},{n_h})]$ computed by the simulations with~\eqref{Pabs1} and (b) with~\eqref{PabsS1} at $2\,{\mathrm{ps}}$ for ${V_{{\mathrm{bias}}}} = 40\,{\mathrm{V}}$. (c) ${n_e}({\bf{r}},t)$ computed by the simulations with~\eqref{Pabs1} and (d) with~\eqref{PabsS1} at $2\,{\mathrm{ps}}$ for ${V_{{\mathrm{bias}}}} = 40\,{\mathrm{V}}$.}
	\label{NeRG3D}
\end{figure}

Fig.~\ref{NeRG3D} (a) and~\ref{NeRG3D} (b) show the spatial distributions of $[G({\bf{E}},{\bf{H}})-R({n_e},{n_h})]$ computed in simulations that use~\eqref{PabsS1} and~\eqref{Pabs1}, respectively, at $2\,{\mathrm{ps}}$ for ${V_{{\mathrm{bias}}}}\! =\! 40\,{\mathrm{V}}$. Figs.~\ref{NeRG3D}~(c) and~\ref{NeRG3D}~(d) illustrate the corresponding ${n_e}({\bf{r}},t)$. In the simulation that uses~\eqref{Pabs1}, the solutions decay smoothly while propagating along the $-z$ direction. This is expected because the optical field is absorbed by the material and screened by the photocarriers. The solutions computed in the simulation that uses~\eqref{PabsS1} are less smooth and ${n_e}({\bf{r}},t)$ is higher near the bottom. Finer meshes are required for stability in the simulation that uses~\eqref{PabsS1}, especially when ${n_e}({\bf{r}},t)$ and ${n_h}({\bf{r}},t)$ are high.

To further show the applicability of the proposed model, the device behavior under different optical pump power levels is studied and compared with experimental results~\cite{Darrow1992}. It is well known that the photocurrent density of photoconductive devices saturates as increasing the optical pump power. This high power saturation behavior results from the coupling between carriers and electromagnetic fields~\cite{Darrow1992, Benicewicz1994, Kim2006, Loata2007, Chou2013, Burford2016, Chen2021screen}. Thus, to be able to model this behavior, the numerical scheme has to take into account the coupling. The first numerical demonstration of this behavior has been given in~\cite{Chen2021screen} using a time-domain multiphysics approach, where excellent agreements between the numerical and experimental data are shown. Here, the same approach is used but the focus is on the comparison of different models used for calculating the carrier generation rate.

\begin{figure}[ht!]
\centering
	\includegraphics[width=0.6\columnwidth]{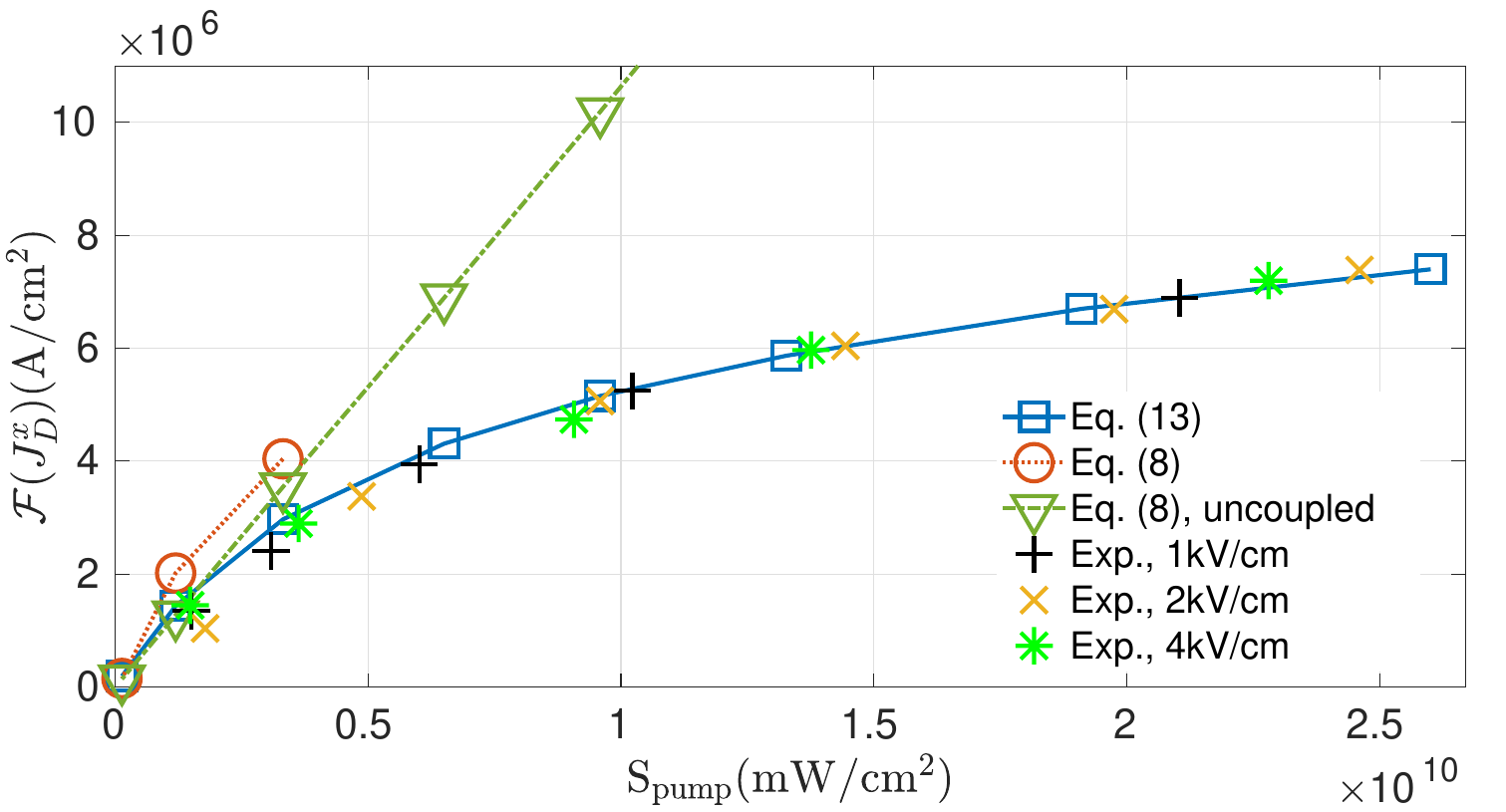}
	\caption{Photocurrent density versus optical pump power simulated using different models and measured in experiments~\cite{Darrow1992}. The experimental data corresponding to different bias fields are scaled appropriately to illustrate the saturation behavior.}
	\label{Saturation}
\end{figure}

In Fig.~\ref{Saturation}, $\mathcal{F}(J_D^x)$ is the $x$-component of the photocurrent density at $1\;\mathrm{THz}$ obtained from Fourier transform of its time-domain signal. The averaged photocurrent density near the top surface (within $100\;\mathrm{nm}$ depth) is used~\cite{Chen2021screen, Yachmenev2019review}. To show more clearly the issue of the Poynting vector-based model caused by the low-frequency radiation, a high bias voltage $V_{\mathrm{bias}} = 40\;\mathrm{V}$ is used to generate stronger low-frequency radiations. Here the experimental data corresponding to different bias fields are scaled appropriately to simply illustrate the saturation behavior (but not for a quantitative comparison). Note that the results for lower bias voltages that correspond to the experimental setup have been given in~\cite{Chen2021screen} and are not represented here. 

Fig.~\ref{Saturation} shows that the saturation behavior modeled with the proposed model~\eqref{Pabs1} matches with the experimental result well. However, when~\eqref{PabsS1} is used, the carrier density diverges when $S_{\mathrm{pump}}$ is large. Hence, only a few data points at low $S_{\mathrm{pump}}$ are obtained. For the case of using~\eqref{PabsS1} but without coupling, $\mathcal{F}(J_D^x)$ increases almost linearly as increasing $S_{\mathrm{pump}}$. Moreover, using~\eqref{PabsS1}, the photocurrent density obtained with coupling is clearly higher than that of the uncoupled case. This is physically unreasonable since the coupling yields various screening effects~\cite{Darrow1992, Benicewicz1994, Kim2006, Loata2007, Chou2013, Burford2016, Chen2021screen} that are expected to weaken the photocurrent.

\section{Conclusion}
\label{concl}
The strong nonlinear coupling between electromagnetic fields and photocarriers in optoelectronic devices calls for a time-domain numerical scheme. An important step in time-domain simulations is the calculation of the carrier generation rate from the electromagnetic fields. The Poynting vector-based model overestimates the carrier generation because the Poynting vector includes the power flux density of the low-frequency fields generated by the photocurrents. This leads to a feedback loop where the overestimated generation rate results in stronger low-frequency fields, which in return increase the generation rate even more. Eventually, solution for carrier densities becomes inaccurate or even divergent when the low-frequency fields are strong.

In this work, an optical absorption-based generation rate model is formulated. First, the optoelectronic material is represented using the Lorentz dispersion model with poles in the optical frequency range. The auxiliary equation, which represents the Lorentz dispersion model in the time domain, is integrated together with the Maxwell and the drift-diffusion equations to update the polarization current density. The polarization current density is used to calculate the instantaneous optical absorption, which is then used to calculate the generation rate. The numerical examples involving photoconductive devices show that the proposed approach is more accurate than the Poynting vector-based scheme and is stable even when the generated low-frequency fields are strong. 

The method developed in this work can be used for time-domain simulations of a wide range of optoelectronic devices, including solar cells, photosensors, and photodetectors. The generation rate corresponding to each Lorentz pole can be calculated independently. This allows for wideband simulations, for example for characterization of photovoltaic devices, to be performed in the time domain using a multipole Lorentz model.

\section*{Acknowledgment}
The authors would like to thank the KAUST Supercomputing Laboratory (KSL) for providing the required computational resources.

\bibliographystyle{IEEEtran}
\bibliography{references_cleaned.bib}

\newpage\clearpage

\section*{Figures}

\newpage\clearpage

\newpage\clearpage


\newpage\clearpage

\newpage\clearpage

\newpage\clearpage

\end{document}